\documentclass[fleqn,usenatbib]{mnras}
\usepackage[utf8]{inputenc}
\usepackage{graphicx}
\usepackage{amssymb}
\usepackage{amsmath}

\usepackage{newtxtext,newtxmath}
\usepackage[T1]{fontenc}
\usepackage{ae,aecompl,ulem,color}
\usepackage[utf8]{inputenc}
\usepackage{graphicx}
\usepackage{amssymb}
\usepackage{amsmath}
\usepackage{xcolor}
\usepackage{natbib}
\usepackage{ulem}
\usepackage{svg}
\usepackage{float}
\usepackage{wasysym}

\newcommand{\Ni}[2]{$^{#1#2}$Ni } 

\newcommand{\solarmass}{\textup{M}_{\odot}} 

\title[Bare Collapse and Fast Optical Transients]{Bare Collapse, Formation of Neutron Star Binaries and Fast Optical Transients}
\author[Mor, Livne \& Piran]{
Ron Mor$^{1}$\thanks{E-mail: Ron.Mor@mail.huji.ac.il}
Eli Livne$^{1}$\thanks{E-mail: Livne@Phys.huji.ac.il}  Tsvi Piran,$^{1}$\thanks{E-mail: Tsvi.Piran@mail.huji.ac.il}
\\
$^{1}$Racah Institute of Physics, The Hebrew University of Jerusalem, Jerusalem, 91904, Israel
}
\date{March 2022}

\begin{document}

\maketitle

\begin{abstract}
{``Bare collapse", the collapse  of a bare  stellar core to a neutron star with a very small mass ejection links two seemingly unrelated phenomena:  the formation of binary neutron star ({BNS}) systems  and the observations of  fast and luminous optical transients. We carried out calculations of the collapse due to electron-capture of both evolutionary  and synthetic isentropic  bare stellar cores. We find that the collapse results in {the formation of} a light $\sim 1.3 \solarmass$ neutron star and {an} ejection of $\sim 0.1 \solarmass$ at $\sim 0.1c$. The outer shell of the ejecta is composed of \Ni56 that can power an ultra-stripped supernova. The models we explored can explain most of the  observed fast optical flares but not the brightest ones.  Collapse of cores surrounded by somewhat more massive envelopes can produce larger amounts of \Ni56 and  explain  brighter flares. Alternatively, those events can arise due to interaction of the very energetic ejecta with winds that were ejected from the progenitor a few days before the collapse. }
\end{abstract}


\section{Introduction}
\label{sec:Introduction}
Numerous evidence accumulated over the years for the existence of two channels for neutron star {(NS)} formation
\citep{podsiadlowski_2004,piran_shaviv_2005,vdh_2007,vdh_2011,beniamini_2016,tauris_2017}. 
The main channel involves the collapse of a massive star whose  envelope is ejected in a powerful supernova  (SN). 
If more than half of the  mass {of a binary system} 
is ejected during the SN of the secondary, as would be the case in the {SN of a massive star with a NS companion}, the binary will be disrupted unless the collapse involves a kick in the right direction.  
The resulting binary systems will have a large eccentricity and a significant proper motion. 
However,  two thirds of  BNS {systems} have low eccentricity and small proper motion~\citep{beniamini_2016}. 
This requires a second formation channel that operates in the majority of {BNS systems}. In this channel,  the second neutron star  forms without a significant mass ejection and with no kick, e.g.\ via an electron-capture SNe \citep{podsiadlowski_2004}. We refer to this channel as ``bare collapse". 

The discovery of  the binary pulsar system PSR J0737-3039  \citep{burgay_2003,lyne_2004} confirmed this picture \citep{piran_shaviv_2005}. The  orbital parameters of the system (small separation, low eccentricity and location in the Galactic plane)  implied that  the  pulsar J0737-3039B had a very small $m < 1.4-1.5 \solarmass $ progenitor and it  was born with very little{, $< 0.1 \solarmass$,} mass ejection  and {with} no kick\footnote{Detailed calculations suggested that the progenitor of pulsar B, just before the collapse, had a mass of {$\approx1.37\solarmass$} surrounded by a tenuous envelope of lighter elements of $0.1$-$0.16\solarmass$  \citep{Dall'Osso_2014}}. Observations of the predicted very small proper motion, $10$ km/sec \citep{kramer_2006,deller_2009}, confirmed this scenario.   
The rate of these events, as inferred from the fraction of binary pulsars from the total number of pulsars suggests  that bare collapses require unique progenitors (occurring for example in a very narrow mass range or specific metallicity) or unique process taking place during the binary evolution prior to the collapse \citep{tauris_2017}.

The very small mass ejection inferred in particular in PSR 0737-3039, but also indirectly in the majority of  Galactic
binary pulsars, suggests that these events can lead to  fast optical transients {powered by \Ni56 radioactive decay}. 
For large brightness and short duration, the requirements are that a sufficiently large fraction of the ejected mass is \Ni56 and {that} the ejection velocity is large enough. 
Such fast optical transients~\cite[see e.g.][]{drout_2014, arcavi_2016,Pursiainen_2018} are characterised by a fast rise time of a few days and a peak luminosity comparable but typically somewhat lower than regular SNe. The fast rise time limits the mass ejection involved and in many cases  the inferred mass is smaller than the amount of \Ni56 needed to explain the luminosity \citep{arcavi_2016}. However, these estimates assume a regular SN behavior.  Bare collapses ejecta might be different.

In this work we simulate  the collapse of a bare stellar core for both  evolutionary   \citep{jones_2013,tauris_2015} and isentropic   initial conditions. 
We use  a modified version of the hydrodynamics code {\sc Vulcan} \citep{livne_1993},  that includes  both  nuclear reaction chain  and neutrino transport. 
We show that these bare collapse produces neutron stars while causing low mass ejection, hence consistent with the formation channel of most BNS systems.
We focus on calculations of the ejected mass and its composition and velocity. Using these results  we estimate the optical transient that arises from the \Ni56\ decay within this ejecta. The paper is structured as follows. We begin with a review of previous work  {in} \S{\ref{section:earlier work}}. This work is divided to two groups. The first  deals mostly with the collapse and the subsequent mass ejection while the second  deals with the nucleosynthesis and the resulting ultra-stripped SN light curve. We continue in \S{\ref{Computational methods}} with a description of our methods. In  \S{\ref{sec:results}} we present our  results concerning the collapse and the nucleosynthesis within the ejecta.  In \S{\ref{sec:observations}} we describe the resulting optical transient and compare it to observations.
We discuss the results in \S{\ref{sec:discussion}}. We conclude and summarize in \S{\ref{sec:conclusions}}. { In Appendix~\ref{appendix:A} we discuss the effect of neutrinos.}

\section{A brief sumamry of earlier work }\label{section:earlier work}
Electron capture supernova ({ECSN}) takes place when electron capture,  i.e.  $e+p\to n+\nu_e$, reduce the electron degeneracy pressure in a  \mbox{O-Ne-Mg} degenerate core of a massive star leading to collapse. In accretion induced collapse (AIC), a   {white dwarf} ({WD}) accretes mass from a  companion star, until it reaches the Chandrasekhar mass and collapses.  Depending on details of the progenitor, AIC  results in either a neutron star and a possible ejection of a small fraction of the star's mass, or in a type Ia supernova. 

From a computational point of view, but not from an astrophysical one, the ECSN of a bare core
in which almost all the stellar envelope was lost via winds or due to interaction with a companion prior to the collapse, 
is very similar to  AIC (when it does not result in a type Ia SN). 
In both cases, a progenitor of approximately Chandrasekhar mass collapses, and once the collapse begins the triggering mechanism is forgotten and the collapse is driven by electron capture at the center. 
The questions concerned with bare collapse deal both with the nature of the progenitor and the collapse process itself. We divide the discussion accordingly.

\subsection{Progenitor models}\label{subsection:progenitors survey}
Determination  the progenitors for bare collapse is a complicated stellar evolution issue. The critical phase occurs at the very last stages of the stars' lifetimes that are evolving at extreme speed. We don't address this question in this work and we use evolutionary  initial configurations  by \cite{jones_2013,tauris_2015} as well as  isentropic   progenitor models. 

\cite{nomoto_1984,nomoto_1987},  evolved helium cores of massive stars in the range of $\sim 8$-$10\solarmass$. 
In all cases, an O-Ne-Mg core was formed. In the $M_{\mathrm{tot}}=10.4\solarmass$ case, the core exceeded the critical mass for neon ignition ($1.37\solarmass$).
However, in the lighter cases, the cores did not reach neon ignition. Instead,  electron capture  took place   and the systems ended up in ECSNe.

\cite{jones_2013} revisited the evolution of $8$-$10\solarmass$ stars. They evolved stars with initial mass $8.2,8.7,8.75,8.8,9.5,12\solarmass$.
Their calculations showed that the $8.2\solarmass$ star evolved into a stable O-Ne-Mg WD, while the $9.5\solarmass$ and $12\solarmass$ stars ended their lives in type II Fe core collapse supernovae (CCSNe).
It was unclear if the $8.7\solarmass$ star evolved into a stable WD or induced an ECSN. However, the $8.75\solarmass$ and $8.8\solarmass$ stars  ended as stripped bare O-Ne-Mg cores and collapsed in ECSN. 
These results suggest that the promising range of single star progenitors for stripped ECSNe is narrow, with initial mass $8.7\solarmass\leq M\leq9.5\solarmass$

\cite{tauris_2015} presented a systematic investigation of the progenitor evolution leading to ultra-stripped supernovae, i.e.\ SNe whose progenitors  are stellar cores with extremely low helium envelope mass $<0.2\solarmass$.  
Progenitors of this kind can exist  in close binaries
in which tidal stripping by the companion star alters the evolution significantly  \citep[see e.g.][]{podsiadlowski_2004}.  The initial masses of the stars considered by \cite{tauris_2015} are, therefore,  significantly smaller than the initial masses  discussed for single stars.
\cite{tauris_2015} evolved systems of a $2.5$-$3.5\solarmass$ He-star with a $1.35\solarmass$ NS companion, with different orbit periods. They found that ECSN only occurred in a limited range of progenitor mass, $M_{\mathrm{He}}=2.60$-$2.95\solarmass$, depending on the orbit's period.

\subsection{Collapse simulations }
\label{subsection:previous simulations}
The collapse of Chandrasekhar mass degenerate cores  was calculated earlier mostly in the context of AIC. However, from a computational point of view, AIC is similar to our scenario. \cite{woosely_baron_1992} considered a progenitor of C-O white dwarf of initial mass $1.1\solarmass$ which was obtained from Nomoto~\cite{nomoto_1986}. The WD  slowly accreted mass up to approximately the Chandrasekhar limit and then collapsed. 
The calculations didn't reach a very late stage in the simulation, hence, they {found} only  a tiny amount of ejected mass ($\approx10^{-4}\solarmass$) during the prompt phase of the collapse, but they estimated that the neutrino-driven wind will eject about $0.01\solarmass$.

Using the same progenitor, \cite{fryer_1999}  found later  mass ejection of $0.1$-$0.3\solarmass$ under various assumptions on input physics: equation of state (EOS), neutrino physics and relativistic effects. They found  that apart from the EOS, non of the parameters significantly changed the amount of ejected mass,   that varied by factors of at most $3$.
As for the effects of the EOS, they showed that changes in the EOS explain the difference between their results (as well as other similar results such as those of  \cite{hillebrandt_1984} and \cite{mayle_1988}) of $\sim 0.1\solarmass$ mass ejection, and the results of \cite{woosely_baron_1992} which used the same progenitor but failed to eject a significant amount of mass by the shock mechanism\footnote{This   specifically refers to  the shock mechanism, which caused only $10^{-4}\solarmass$ ejected.}

Most recently,  \cite{sharon_kushnir_2020}   used a modified version of the {\sc Vulcan} 1D code \cite{livne_1993} to calculate AIC,  of a synthetic  Chandrasekhar-mass star, with an isentropic core in a hydrostatic equilibrium,
focusing on an accurate treatment of the EOS. 
They found an ejected mass of a few $\times 10^{-2}\solarmass$
with an outflow velocity of  $0.15$-$0.3c$, and the ejecta was composed mainly of \Ni56.  With no neutrino transport the initial $Y_e=0.5$ was kept throughout their calculations, and as we will see later this results in all the ejecta being \Ni56.  

As we  focus on mass ejection it is worth noting that these earlier studies revealed  three main mechanisms for mass ejection in ECSN and AIC \citep{fryer_1999}.
First, in the \textit{prompt} mechanism,  the converging collapse shock bounces off the core, and the resulting diverging shock ejects the outer shells.
Second, in the \textit{delayed-neutrino} mechanism, the bounced shock initially stalls. However, after  $20$-{ $200$m}s, the shock  revives due to neutrino heating and drives  mass ejection. 
Lastly, in the \textit{neutrino driven wind} mechanism, 
neutrino emission by the 
the newly formed hot proto-neutron star is   absorbed in the outer (less dense) layers of the star, causing   mass ejection. 

\subsection{Nucleosynthesis and light curve calculations}
{A different route to understanding ultra-stripped SNe is to investigate the long-term  expansion of the ejecta and compute the resulting light curve. Such works do not follow the dynamics of the progenitor through its collapse. Instead, in such works a large amount of energy is injected to the outer shell of the progenitor, which then drives the mass ejection. The detailed nucleosynthesis and the light curve are computed in post-processing methods after the hydrodynamic calculation. This method is often referred to as explosive nucleosynthesis simulations.}

\cite{moriya_2017} have used this method to compute the nucleosynthesis and light curve by the collapse of an ultra-stripped progenitor computed by \cite{tauris_2013}. The progenitor was initially a $2.9\solarmass$ He star, evolved as a binary of a $1.35\solarmass$ NS. \cite{moriya_2017} found that \Ni56 of mass $\approx 0.03\solarmass$ was formed in the ejecta, and that the rise time of the bolometric light curve was $5$-$10$ days.
The progenitor they used is similar to one of the progenitors we use by \cite{tauris_2015}.

Most recently, \cite{sawada_2021} have performed explosive nucleosynthesis simulations of C-O progenitors of core mass $1.45$-$2\solarmass$ computed by \cite{suwa_2015}. For the lighter progenitors, they found that $0.01$-$0.02\solarmass$ of \Ni56 was synthesized in the ejecta, and the light curves had rise times of a few days.

\section{Methods}
\label{Computational methods}

\subsection{The overall scheme} 
We use the {\sc Vulcan} code \citep{livne_1993},
with some  modifications, 
to carry out one-dimensional simulations, including hydrodynamics, neutrino transport and nuclear burning. 
The hydrodynamics is non-relativistic, and its scheme is explicit and Lagrangian.
We  used an adaptive mesh refinement~(AMR) mechanism to allow a dynamical refinement of the mesh in the important regions. 
AMR was mainly used to decrease the resolution of the newly formed NS after bounce, and to increase the resolution in the ejected mass at late times.   
The initial grid in our standard simulations consists of 2842 cells, about half of them describe the inner $\sim 1.2\solarmass$ which eventually remains bound, and the remaining cells describe the outer envelop. In this initial grid, the cell mass was equal to $10^{-3}\solarmass$ for most of the star, where the cells at the outer region of the progenitor were increasingly smaller in mass, as low as $5\times 10^{-5}\solarmass$.
At later times, once the PNS has formed and no longer affects the ejecta, we gradually reduce the resolution of it to $\sim 100$ cells. 
{We checked convergence by reducing the resolution with respect to two main parameters, first increasing the size of the numerical cells in the initial grid, and secondly increasing the maximal allowed size of a numerical cell in the AMR mechanism. Reducing the resolution by a factor of few with respect to these parameters did not change our qualitative results. For the isentropic   progenitor (see~\S\ref{section:isentropic stars}), in either resolution a PNS was formed, with ejection of $\sim 0.1\solarmass$, with similar composition. Quantitatively, the results varied by up to $10$-$20$\%.}

\subsection{The Equation of state}\label{subsec:eos} 
The densities in the  collapse  vary from\footnote{At late times of the expansion of the ejecta, even lower densities are obtained.} $\sim 10^{5}{\mathrm{g}}~{\mathrm{cm}^{-3}}$ to $\gtrsim 10^{14}{\mathrm{g}}~{\mathrm{cm}^{-3}}$.  We use two different EOSs for the different thermodynamical regimes.
For high densities, namely the part of the progenitor which eventually results as part of the NS, we use a tabulated EOS for nuclear material. 
The EOS is based on tables originally provided by \cite{shen_1998_1,shen_1998_2} \footnote{This EOS was also used in \cite{sharon_kushnir_2020}.}.
The EOS table (compiled by \cite{oconnor_2010})
utilizes the relativistic mean field (RMF) theory, and calculates the EOS for homogeneous nuclear matter, as well as for inhomogeneous matter using the Thomas-Fermi approximation. The matter is assumed to be in NSE, and to be comprised of a mixture of neutrons, protons, alpha-particles, a single species of heavy nuclei, and leptons.
For the lower densities, namely the part of the progenitor which is later ejected, we use the EOS of degenerate electrons gas with free ions.

\subsection{Neutrino transport} 
The neutrino transport scheme (based on an unpublished work of   Eli Livne) solves the transport equation implicitly in the co-moving frame, adequately treating transparent regions and opaque ones.  
The scheme is inherently built to approach the diffusion method for opaque regions, and the discrete-ordinates method $S_n$ for transparent regions. For the $S_n$ method, we usually used $5$ azimuthal partitions, i.e.\  $n=5$, but we checked $n>5$ as well and found no significant differences in the results.
The neutrinos cross-sections were taken from \cite{burrows_2006}. We used $18$ energy groups for the neutrinos, {with the bin cetners} between {$0.5$-$286.3$} MeV, with a logarithmic spacing of the {size of the} energy bins. The size of the $k$'th energy bin {is}  $\approx 1.29^{k-1}$ MeV. 

In our simulations, we considered  only $\nu_e$  and $\bar \nu_e$. 
We estimate that the effect of the $\nu_\nu$ and $\nu_\tau$ flavors to be insignificant for the results we are interested in. These neutrinos are not involved in nuclear interactions  that change the electron fraction.  Furthermore, their  absorption cross sections are weaker, reducing their impact on the ejecta. We expect that the main effect of these neutrinos would be enhancing the cooling rate in the late phase of the process, once the neutron star forms. 

\subsection{Nuclear chain}
Nuclear burning was calculated using reactions rates based on \cite{rauscher_2000,rauscher_2001} with a network of $54$ isotopes.
We validated the nuclear burning results using {\sc SkyNet} \citep{lippuner_2017_skynet} as discussed in \S\ref{subsection:isen ejected mass}.

\subsection{The progenitors }
\label{subsection:our progenitors}
Numerous  possible progenitors may evolve in different stellar evolution environments, during the last period of star's lifetime
as discussed in \S\ref{subsection:progenitors survey}. We do not solve this question here, but rather consider two different types of progenitors, namely synthetic isentropic progenitors, and   evolutionary   progenitors obtained from  detailed stellar evolution studies. We show that {the results for both types are similar. This allows us to use the  isentropic   progenitors as a generic model}.

\begin{figure} 
 \begin{center}
\includegraphics[scale=0.57]{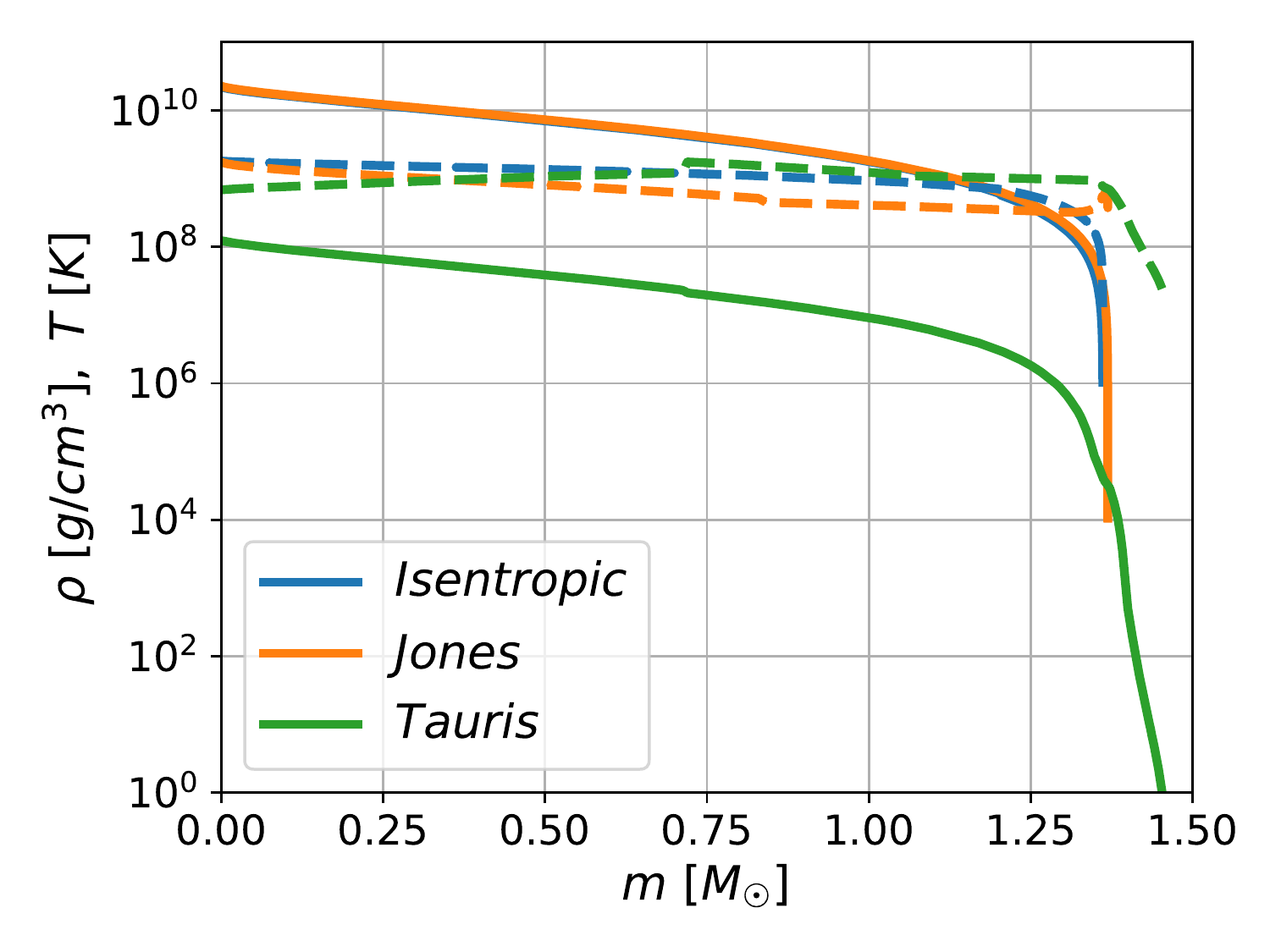}
\vskip-0.3cm
\caption{Initial density (full lines) and temperature (dashed lines) for the  isentropic   progenitor (blue), and for the  evolutionary  progenitors of \protect\cite{jones_2013} (orange) and  \protect\cite{tauris_2015} (green). The density profiles for the isentropic progenitor and for the progenitor of \protect\cite{jones_2013} are almost identical.}
\label{fig:progenitors}
\end{center}
\end{figure}

Our first type of progenitor is  isentropic Chandrasekhar-mass white dwarfs in hydrostatic equilibrium.  For our standard simulations we used a progenitor with a central density of $2.2\times 10^{10}{\mathrm{g}}~{\mathrm{cm}^{-3}}$ and a central temperature of $1.8\times 10^{9} \mathrm{K}$. Fig. \ref{fig:progenitors} shows the thermodynamic properties of the progenitor.

A second  progenitor was evolved by \cite{jones_2013} from  He core of a $8.75\solarmass$ star   (see  \S\ref{subsection:progenitors survey}). 
It is a degenerate core of mass $1.37\solarmass$, most of it composed mainly of $^{16}$O and $^{20}$Ne, where the outer $0.02\solarmass$ is composed mainly of $^{12}$C with some $^{16}$O, $^{20}$Ne and $^{24}$Mg.
Fig. \ref{fig:progenitors} shows the initial profile. It has a central density of $2.27\times 10^{10}{\mathrm{g}}~{\mathrm{cm}^{-3}}$ and a central temperature of $1.75\times 10^{9}\mathrm{K}$, similar to our standard isentropic   star.  
At its initial state, the progenitor
is at the onset of collapse 
after electron capture has very mildly started  at its center. The initial electron fraction of the   progenitor  is slightly smaller than $0.5$ for most of the star.
The material has  small inwards velocities, but the kinetic energy is negligible compared to the gravitational energy of the resulting NS. As a  test, we performed minor adjustments to the initial profile so it would be in hydrostatic equilibrium. The results of the collapse of this adjusted profile were very similar to the results of the original profile, and  we do not discuss it further in this work.

We also consider  a third progenitor calculated by  \cite{tauris_2015}. This progenitor arises as a result of a binary evolution, as discussed in~\S\ref{subsection:progenitors survey}. It is considerably less dense compared to the other progenitors we examined, with the central density  lower by a factor of $\sim 100$ than the central density of  the progenitor found  by \cite{jones_2013} for the single evolution. 
Figure \ref{fig:progenitors} 
shows the density and temperature profiles of the main three progenitors we used in this work.

\section{Results}
\label{sec:results}

\subsection{Isentropic  progenitor}
\label{section:isentropic stars}
We discuss the main results of our simulation for the isentropic progenitor.
The star is initially in hydrostatic equilibrium. As electron captures occurs {spontaneously} at the center,  a region of low $Y_e$ expands from the center of the star outwards, and the value of $Y_e$ decreases in time as electron capture keeps occurring. This causes an instantaneous reduction of the pressure due to the  removal of degenerate electrons.  This process continues and induces the collapse of material towards the center of the star.
Eventually, material bounces back from the center {as nuclear densities are achieved}, launching  a diverging shock wave that ejects the outermost layers. 

\begin{figure} 
\begin{center}
\includegraphics[scale=0.34]{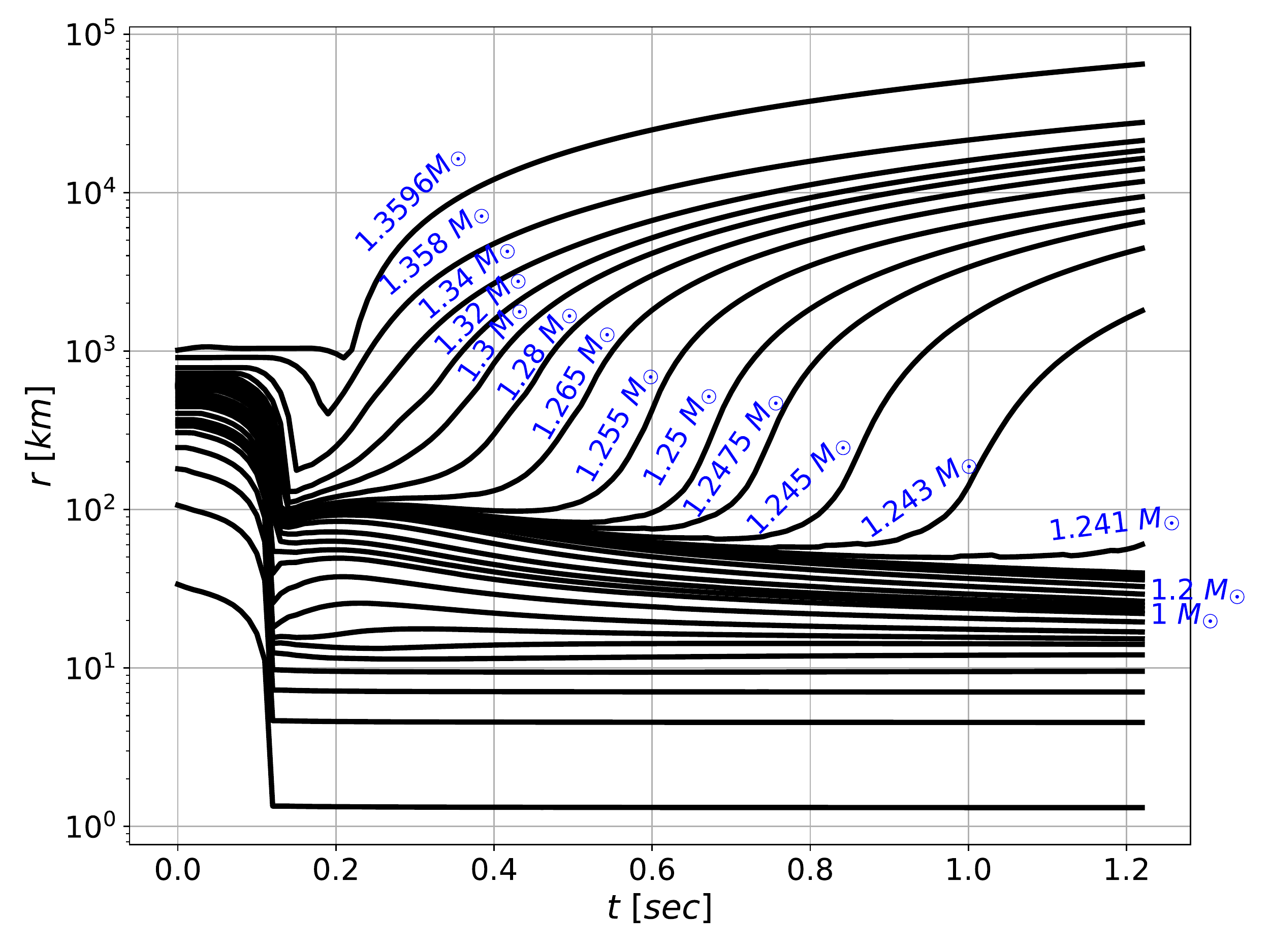}
\caption{Trajectories of  mass elements for the isentropic   progenitor. Bounce and shock breakout occur at  $t \approx 0.12$ sec and $t\approx 0.2$ sec, respectively. Approximately $0.08-0.1\solarmass$ is ejected off the star immediately, and additional $0.02-0.04 \solarmass$ is ejected over $\approx 1\mathrm{\ sec}$.}
\label{fig:isen rtplot}
\end{center}
\end{figure}

Fig.  \ref{fig:isen rtplot}  depicts  mass elements trajectories as a function of time. We see that the star starts its collapse immediately. At  $t \approx 0.12$ sec the bounce occurs and the diverging shock wave forms.  Shock breakout occurs at  $t \approx 0.2$ sec and the outermost  layer of the star is ejected. The  remnant  is a compact star of mass $\approx 1.24\solarmass$ which is initially at radius $\approx 40$ km. It is a proto-neutron star (PNS) which is still very hot, and so has a rather large radius. It takes for the PNS tens of seconds to achieve a standard NS radius, and we will discuss the late time evolution of the PNS in~\S\ref{sec:neutron-star}.
 
\subsubsection{The ejected mass}\label{subsection:isen ejected mass}
\begin{figure}
\begin{center}
\includegraphics[scale=0.57]{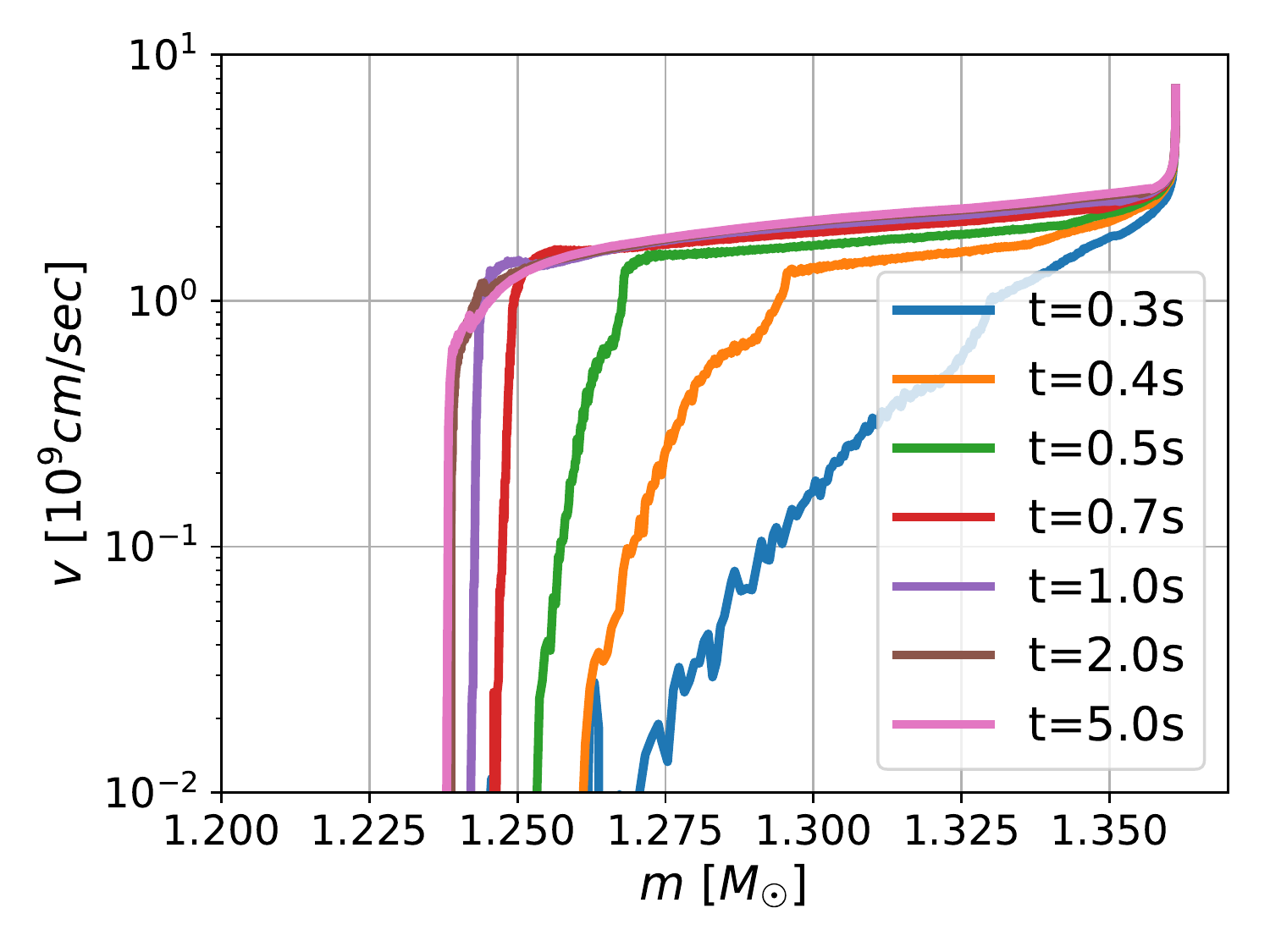}
\vskip-0.5cm
\caption{Velocity vs mass coordinate at different times. The ejecta accelerates during the first $\approx 0.7$ sec and then expands homologously, with velocities between $0.05c$ and $0.1c$ for most of the ejecta.}
\label{fig:isen v_vs_m}
\end{center}
\end{figure}

About $0.08-0.1 \solarmass$ is ejected immediately due to the shock, at  $t \approx 0.2$ sec. 
Later on, a smaller amount of $0.02-0.04 \solarmass$ is ejected  over a period of $\approx 1$ sec in a decreasing rate\footnote{Additional $3\times 10^{-3}\solarmass$ are ejected at later times, mostly around  $\sim 2$ sec, (not shown in Figure~\ref{fig:isen rtplot}) due to {additional} neutrino driven wind.}. 
We attribute this ejection to neutrino absorption and the neutrino driven wind mechanism. 
Fig.  \ref{fig:isen v_vs_m} shows the velocity profiles of the ejected mass at different times. We see that the ejected mass accelerates up to time $\approx 0.7$ sec, and then expands homologously. Most of the ejected mass travels at a small range of velocities, starting from $1.5\times 10^{9}{\mathrm{cm}}~{\mathrm{sec}^{-1}}$ at the inner regions of the ejecta, up to $3\times 10^{9}{\mathrm{cm}}~{\mathrm{sec}^{-1}}$ for {the} outermost  regions. A tiny amount of mass (about $10^{-3}\solarmass$) expands at up to about $7\times 10^{9}{\mathrm{cm}}~{\mathrm{sec}^{-1}}$.

The  ejecta contains at  the time of shock breakout, $t\approx 0.2$ sec,  only  neutrons, protons and alpha particles. 
The temperatures behind the shock, at this time, are between $\approx 3$~MeV for the inner parts of the ejecta and $\approx 1$ MeV for the outer parts. 
The temperatures are so high that nuclei heavier than $^{4}$He disintegrate. 
The densities of the inner regions of the ejecta are at this time, a few times $10^{10}{\mathrm{gr}}~{\mathrm{cm}^{-3}}$, 
and these regions are comprised  mostly of neutrons. 
Upon shock breakout, the material expands and cools. Starting from the outermost  layer, and proceeding inwards as time progresses, heavier elements, in particular $^{56}\mathrm{Ni}$, form as the ejecta expands and so its density and temperature decrease. This stage occurs at densities of $\sim 10^{7}\mathrm{g}/\mathrm{cm}^3$ and temperatures of $\sim 0.5$~MeV, slightly larger for the inner ejecta compared to the outer ejecta. {For each mass element, the burning phase is very short.} The composition freezes out once the temperature and density drop slightly below this range. 

\begin{figure}
\begin{center}
\includegraphics[scale=0.57]{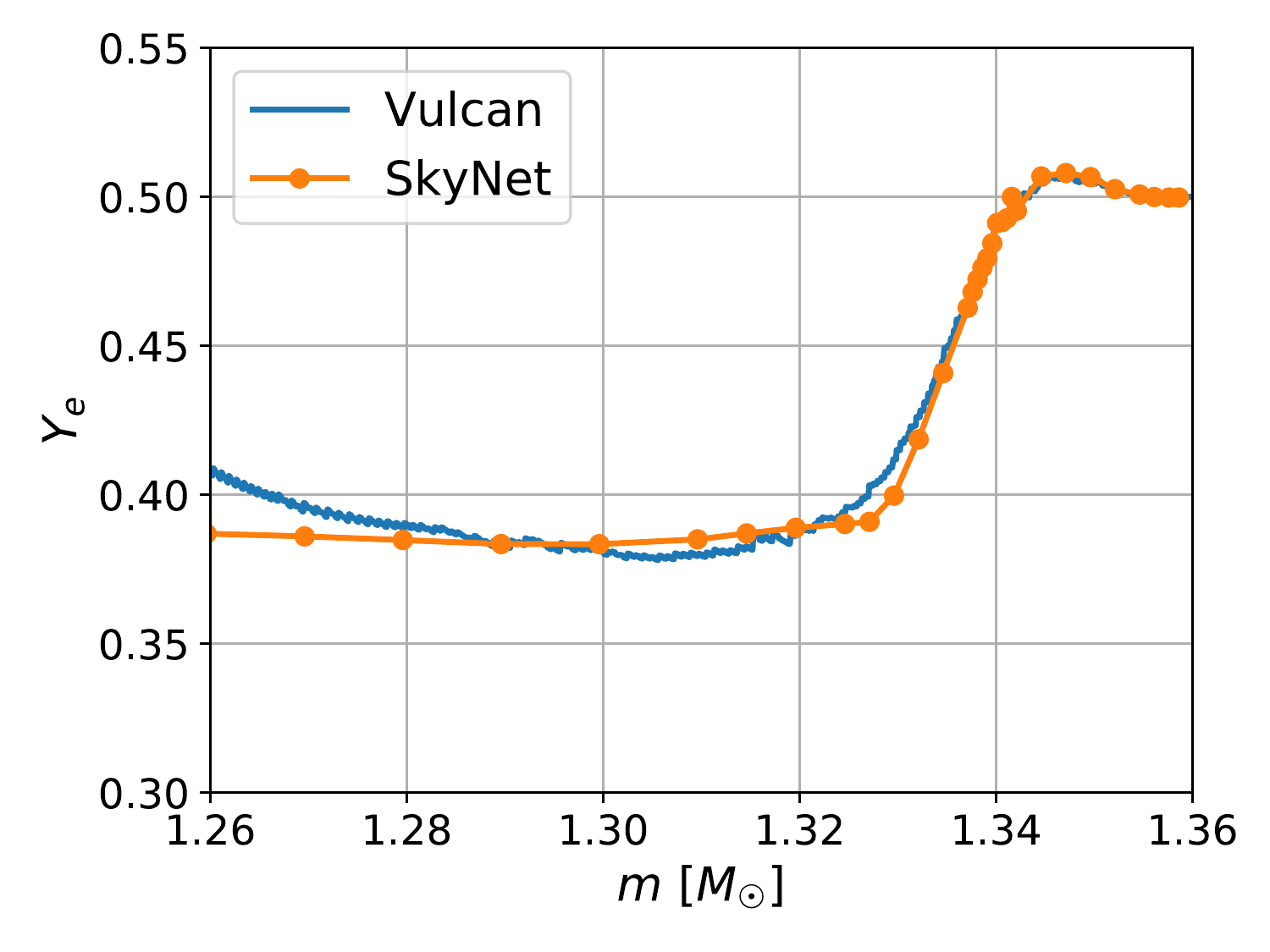}
\caption{Electron fraction profile of the outer $0.1 \solarmass$ of the ejecta, by the {\sc Vulcan} simulation (blue) {and the} {\sc SkyNet} simulations (orange). The match is overall excellent, and the region of $Y_e\approx 0.5$ corresponds to the region where \Ni56 is formed.  
}
\label{fig:skynet_ye}
    \end{center}
\end{figure}

\begin{figure}
\begin{center}
\includegraphics[scale=0.57]{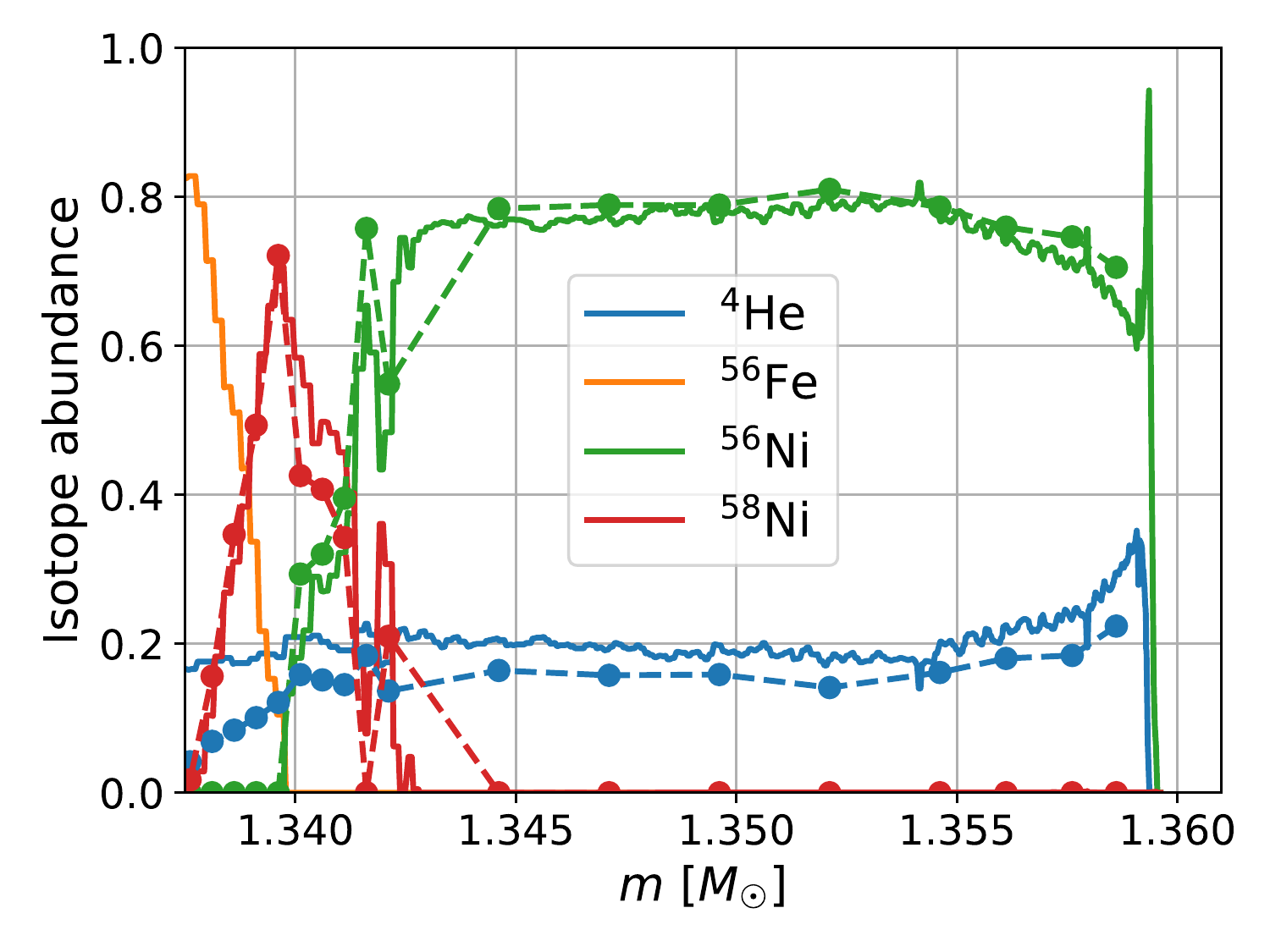}
\vskip-0.3cm
\caption{ 
The abundance of $^{56}$Ni,$^{58}$Ni,$^{4}$He and $^{56}$Fe for the outer $0.225\solarmass$ of the ejecta, by our full {\sc Vulcan} simulation (full lines) and by {\sc SkyNet} (dashed lines). The results are almost identical for the $^{56}$Ni,$^{58}$Ni and $^{4}$He abundance. $^{56}$Fe that appears in the {\sc Vulcan} results does not form in the {\sc SkyNet} simulations in which the lower-$Y_e$ region comprises of many different iron peak isotopes that are not part of our standard {\sc Vulcan} nuclear network.   
}
\label{fig:skynet_compos_compare}
\end{center}
\end{figure}

At  $t\approx 0.6$ sec the composition of all of the ejecta freezes. 
The outer $\approx 0.02\solarmass$ of the ejecta has $Y_e\approx 0.5$ (see Fig.~\ref{fig:skynet_ye}). This region has much lower density than the rest of the ejecta. Therefore,  electron capture is substantially lower and $Y_e$ does not change from its initial $0.5$ value. 
$^{56}$Ni  is the most abundant isotope in NSE composition of relatively dense matter with $Y_e=0.5$ (partially because its nucleus has the same number of protons and neutrons). Hence $^{56}$Ni forms in this outermost region. 
$^{56}$Ni was created only at this very outer shell of the ejecta, which has important observational implications (see  \S\ref{sec:observations}). 
The total \Ni56 mass was $\approx 0.02\solarmass$.
The composition of the ejected mass is shown in Fig. \ref{fig:skynet_compos_compare}, where we focus on the outermost part of the ejecta, which is the region that \Ni56 was formed at and that was validated using {\sc SkyNet} (see later).
In a small inner shell we see a peak of $^{58}$Ni.
The rest of the ejecta, which is most of it, has \mbox{$Y_e\approx 0.4$} (see Figure~\ref{fig:skynet_ye}).
In our {\sc Vulcan} simulations $^{56}$Fe is created  in this  low $Y_e$ region, but  detailed nucleosynthesis calculations using {\sc SkyNet} (see later) show that the composition in this region is more complicated and comprises of many iron group isotopes. 

{\sc Vulcan} employs a rather large network of 54 isotopes. However, it is clearly somewhat limited.  We validated the {\sc Vulcan} results using {\sc SkyNet} ~\citep{lippuner_2017_skynet}, which includes 7836 isotopes and about 93000 reactions. 
We used the thermodynamic trajectories of different mass elements from the {\sc Vulcan} simulation as an input to {\sc SkyNet}, which solved the nuclear reaction equations. 
To account for the neutrino history of the mass elements, we  assumed that the neutrinos were emitted at a time dependent rate from a source at the origin, as determined by the neutrino luminosity computed in our {\sc Vulcan} simulations. The source 
is specified by a simplistic assumption of a Fermi-Dirac distribution with zero chemical potential, with an average energy of $k_B T_{\mathrm{source}}$, where the temperature determining the distribution was taken to be time dependent as well and is defined as follows.
At the early stages after the collapse of up to $\lesssim 1$ sec, when the nucleosynthesis in the ejecta occurs, the PNS is qualitatively divided to two regions - an inner dense core of densities over $10^{14}{\mathrm{g}}/{\mathrm{cm}^3}$, and an outer envelope of much smaller densities that still accretes  onto the  PNS (see Figure~\ref{fig:isen PNS}). We assume that the inner dense core does not contribute to the neutrino flux at these early times as it is opaque, while the outer region is  transparent. Therefore, we average the temperature only over the outer region with densities $<10^{14}{\mathrm{g}}/{\mathrm{cm}^3}$, and define the time dependent source temperature 
by $T_{\mathrm{source}}^4(t)=\int_{\mathrm{outer\ NS}}T^{4}(t,m)dm/\int_{\mathrm{outer\ NS}} dm$
with respect to the accumulating mass coordinate.
This assumption mimics the presence of the remnant PNS, which is the source of neutrinos during the time that nucleosynthesis takes place.

We obtained an  overall excellent match of $Y_e$ between our  {\sc Vulcan} simulation and the {\sc SkyNet} simulation, as shown in Figure~\ref{fig:skynet_ye}. Fig.  \ref{fig:skynet_compos_compare} shows a comparison of the mass fraction of $^{56}$Ni,$^{58}$Ni,$^{4}$He and $^{56}$Fe (the main isotopes that were obtained in the {\sc Vulcan} simulations) in the outer $0.0225 \solarmass$ of the ejecta, {as calculated} by the  {\sc Vulcan}  and  the {\sc SkyNet} simulations. 
{\sc SkyNet} produces a similar composition in this outermost region, with $^{56}$Ni being the dominant isotope, confirming {\sc Vulcan}'s results. 
In both {\sc Vulcan} and {\sc SkyNet} simulations, the region with $Y_e\approx0.5$ is precisely where $^{56}$Ni is formed\footnote{Since $Z/A$ for $^{56}$Ni is $0.5$ it is the most abundant isotope in fluid elements where $Y_e\approx 0.5$. However, in places where $Y_e$ is smaller, $^{56}$Ni is barely present if at all.}, and the two programs agree on the size of this region and its the composition, so correspondingly the total amount of \Ni56 in the ejecta is the same.


\subsubsection{The Neutron Star}
\label{sec:neutron-star}
The relevant time scale for the PNS to achieve equilibrium is much longer than the typical time scale for  the ejection of the outer stellar layers  or for the nucleosynthesis in the ejecta.
To explore the later  evolution  as the PNS  turns into a standard NS, we ran a longer simulation with two main changes. 
First, the ejecta was removed  at $t= 0.4$ sec. This  is late enough for the dynamics of shock breakout and the mass ejection to have occurred, so the PNS and the ejecta are  largely decoupled\footnote{Late neutrino flux influences the ejecta but not the PNS.} at this stage and we may focus only on the PNS. 
Second, the resolution was reduced to allow simulating long physical times with a reasonable run time. In this section we show results from our standard simulation up to $t=5$~sec, and from this modified simulation for $5<t\lesssim 30$~sec. 

\begin{figure}
\begin{center}
\includegraphics[scale=0.57]{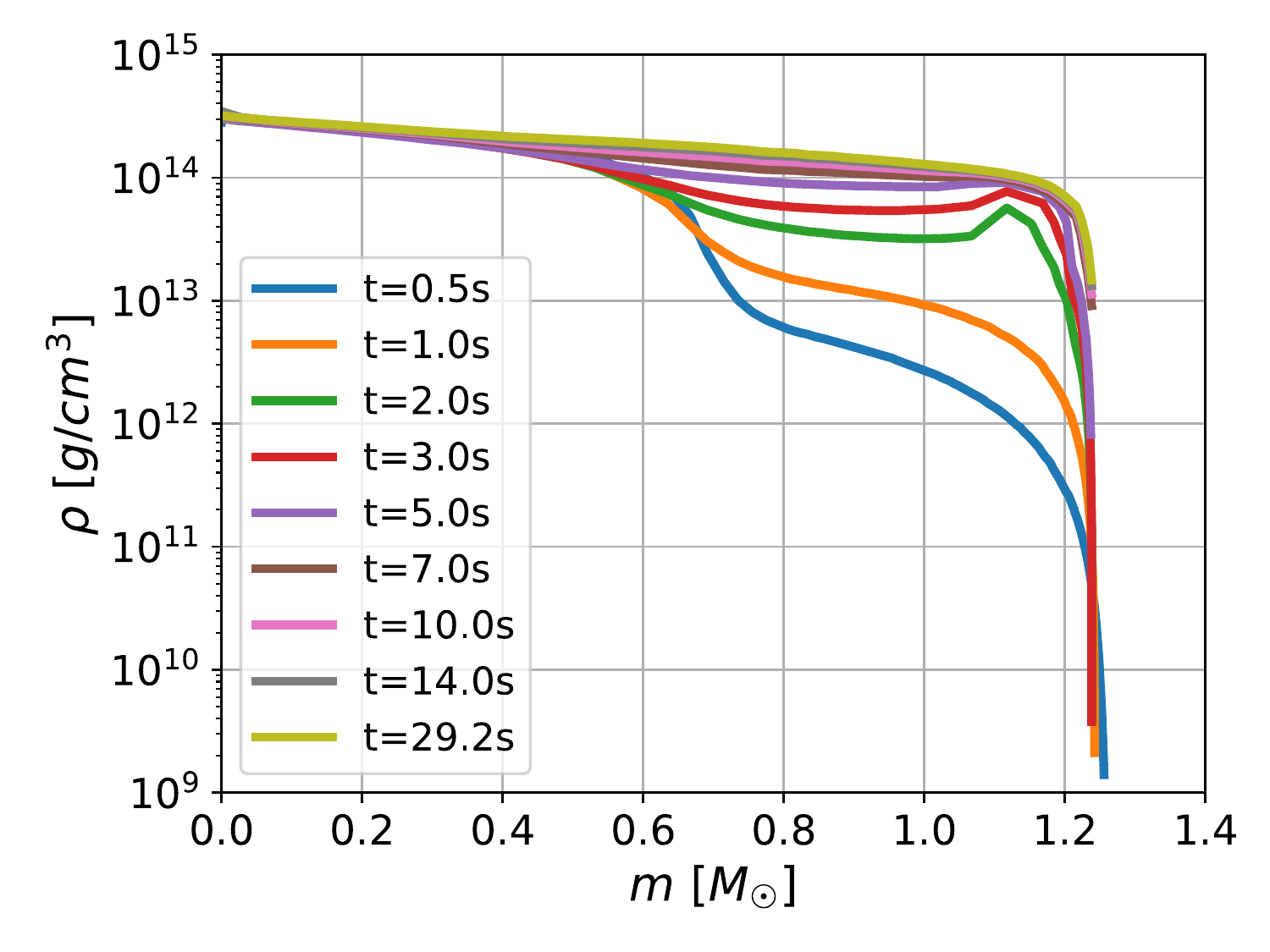}
\vskip-0.cm
\includegraphics[scale=0.57]{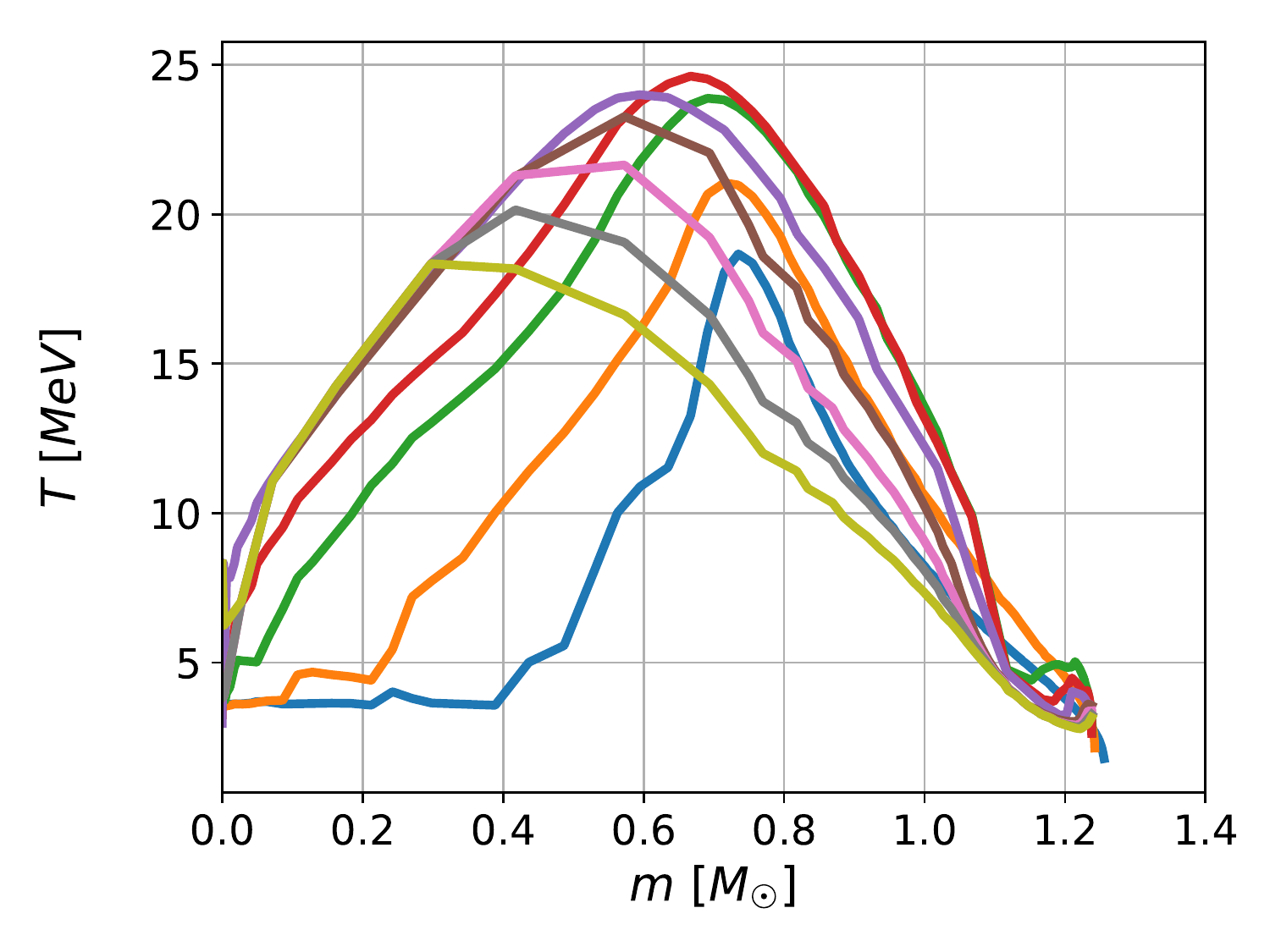}
\vskip-0.3cm
\caption{
Density (top) and temperature (bottom) vs mass for the PNS, at different times. 
It takes a few seconds for the outer region ($\approx 0.5\solarmass$) of the PNS to accrete and reach nuclear densities.
At the first few seconds, the temperature increases  as the PNS evolves and contracts. Then, the PNS heats from the middle towards the center, while cooling off its outer boundary.
The profiles up to time $5$s were taken from our standard simulation. Later profiles  were taken from a similar simulation tailored to allow simulating long physical times with a reasonable accuracy (see text). }
\label{fig:isen PNS}
\end{center}
\end{figure}

Figure~\ref{fig:isen PNS} depicts the density and temperature profiles of the PNS for  different times. 
The inner $\approx 0.7$-$0.8\solarmass$ approaches nuclear density almost immediately. 
This critical point of $m\approx 0.8\solarmass$ is where bounce occurred during the collapse
at approximately $120$ msec. 
The outer part of the PNS is initially at densities lower by $1$-$2$ orders of magnitude. 
Even {after $\sim 1$} sec there are still some dynamics at this region and matter keeps falling towards the PNS and its density increases. The temperatures range from $3$ MeV to $22$ MeV peaking  at $m\approx 0.7\solarmass$-$0.8\solarmass$, where the bounce took place and where we saw a qualitative change in behaviour of the density profile. 
The temperatures  keep increasing for a while as  gravitational energy is released due to the contraction of the PNS that is turning into an ordinary NS.

\begin{figure}
\begin{center}
\includegraphics[scale=0.57]{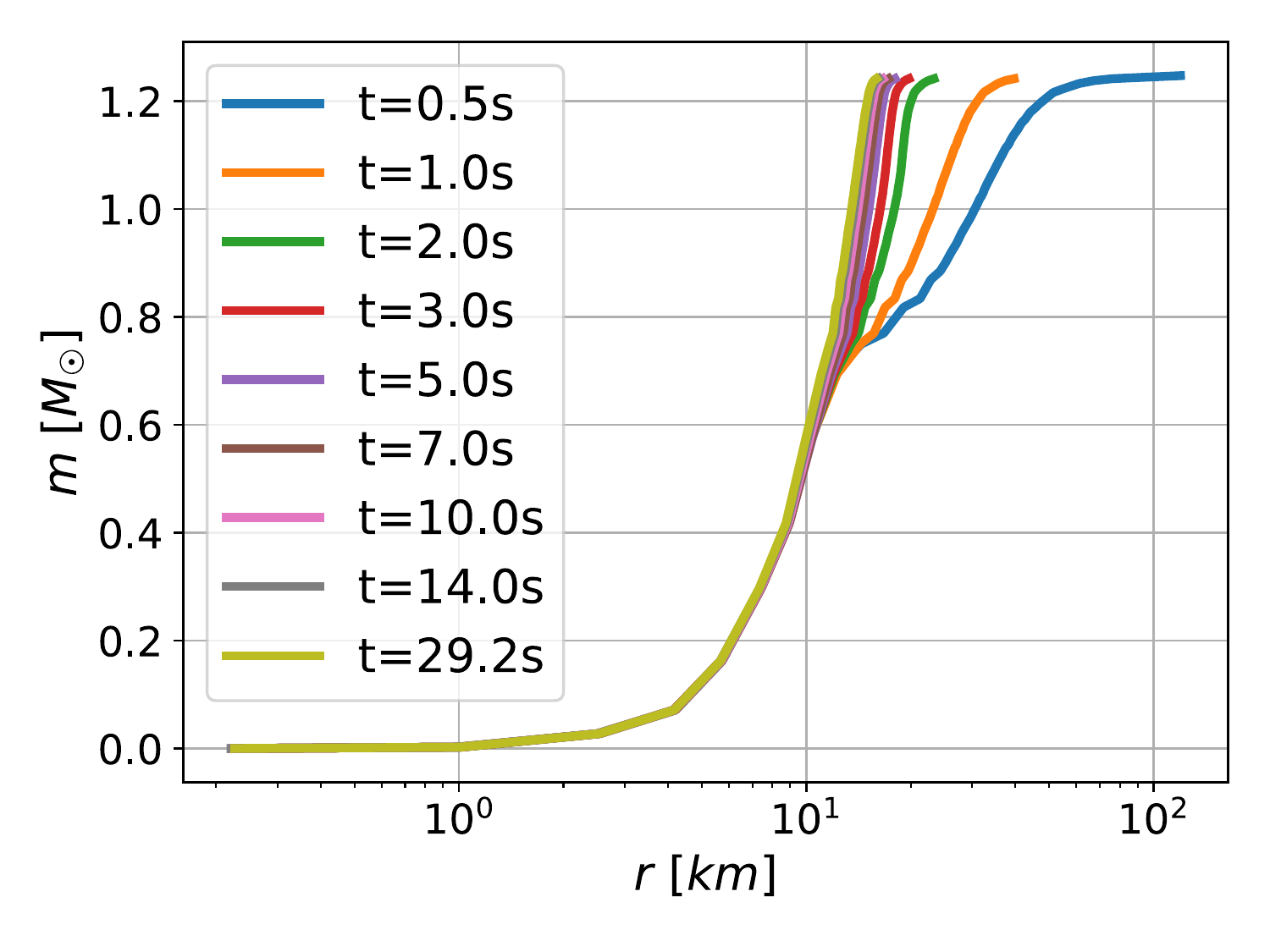}
\vskip-0.3cm
\caption{Accumulating mass vs radius, at different times. It takes many seconds for the PNS to reach an ordinary NS radius. The profiles at all times are taken from our  lower-resolution simulation (see text), but the results are the same as in our standard simulation.}
\label{fig:isen m_vs_r_PNS}
\end{center}
\end{figure}

The dynamics of the PNS keep going for a long period. The  radius of the PNS, that was approximately $40$ km at  $t\approx 1$ sec, decreases  to  $\lesssim 15$ km at the latest time of our simulation  $t\approx 30$ sec{, see Figure~\ref{fig:isen m_vs_r_PNS}}. 
The general picture at these late times, which we explain shortly, is similar to the known theory of PNS \citep[see e.g.][]{burrows_1986}.
Neutrinos escaping from the outer shells of the PNS reduce the neutrino radiation pressure and allow the outermost  layers of the PNS to  accrete on the inner PNS core. This causes the increase in temperature at early times.
Next, the hot and opaque PNS heats inwards, while electron capture continuously reduces the electron fraction, until a hot NS is formed.
Finally, after tens of seconds have passed, the hot NS cools down by neutrino emission  and the NS becomes  more compact. 
Note that our simulations included only $\nu_e$ and $\bar \nu_e$. Inclusion of the $\nu_\mu$ and $\nu_\tau$ flavors could accelerate the cooling process (but won't change the qualitative behaviour).

\subsection{ An evolutionary  single-star progenitor}\label{section:jones}

Next we simulated the collapse of an evolutionary progenitor calculated by  \cite{jones_2013}. 
The collapse of this progenitor resulted in {the formation of a} NS and a small amount of mass ejection, qualitatively and quantitatively very similar to the results discussed in \S\ref{section:isentropic stars}.
The velocity, mass, composition and electron fraction of the ejecta are very similar for both types of progenitors.
This similarity is reassuring and shows the robustness of our key result, which is an ejected shell of $\approx0.13\solarmass$ traveling at $0.1 c$, where the outer $0.02 \solarmass$ shell is composed of $77\%$ $^{56}$Ni.

\begin{figure} 
\begin{center}
\includegraphics[scale=0.34]{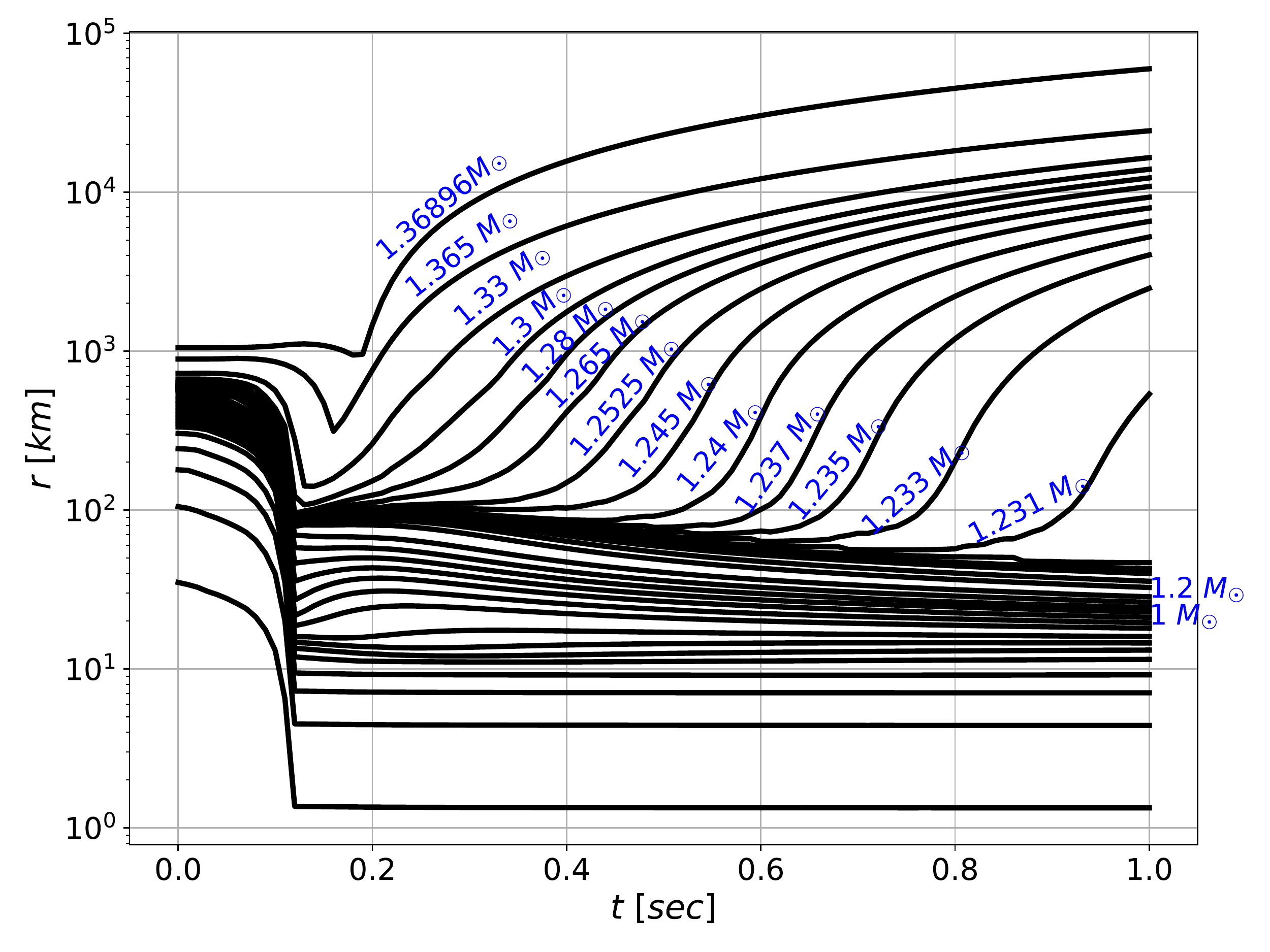}
\caption{Trajectories of  mass elements for the evolutionary  single-star progenitor {of~\protect\cite{jones_2013}}. Approximately $0.14\solarmass$ is ejected due to the gravitational collapse, leaving a PNS of mass $1.23\solarmass$. The results are very similar to the case of the isentropic   progenitor.}
\label{fig:Jones rtplot}
\end{center}
\end{figure}

\begin{figure}
\begin{center}
\includegraphics[scale=0.57]{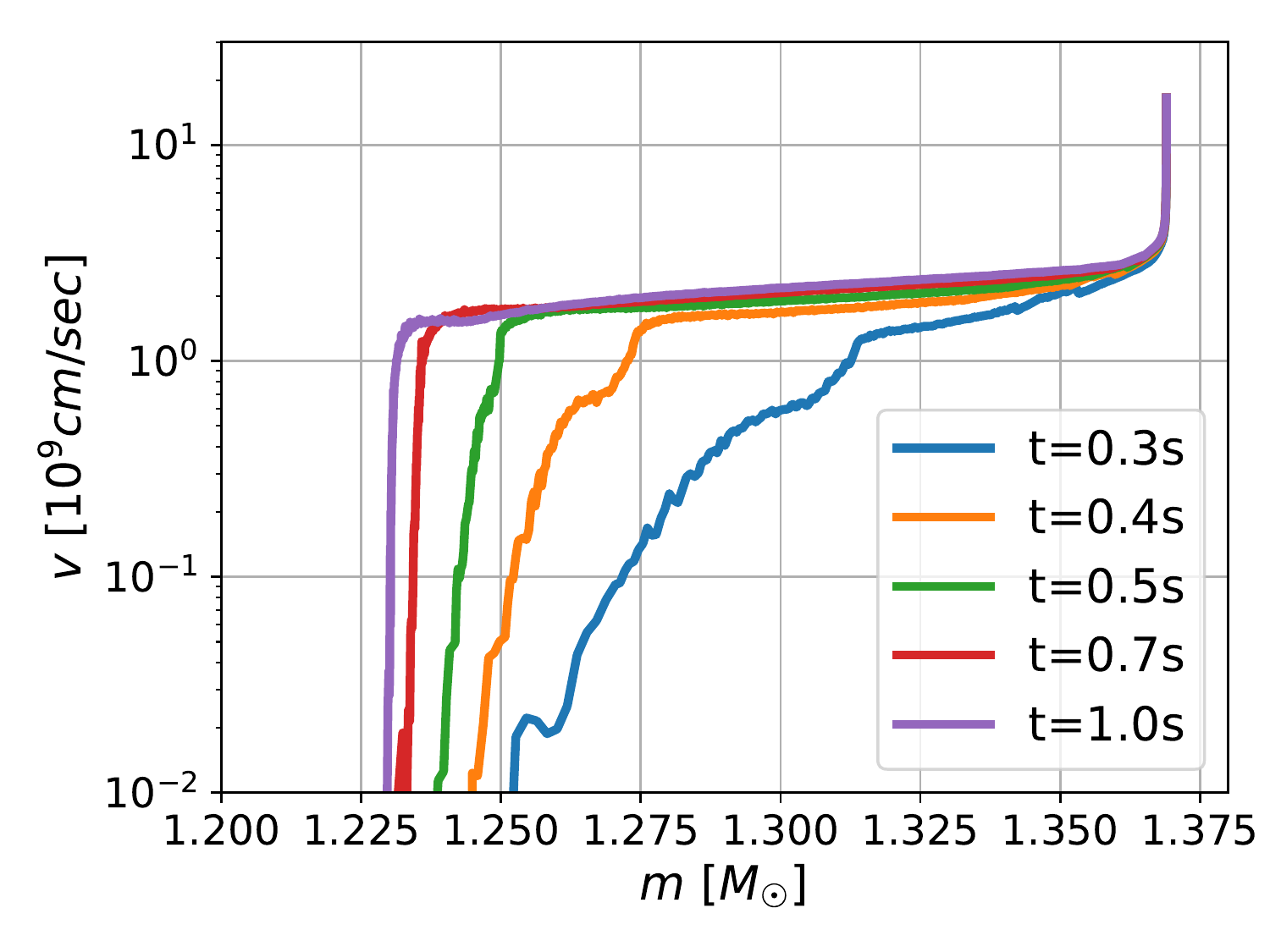}
\vskip-0.3cm
\caption{Velocity profiles  of the {outermost region of the} ejecta at different times, for the {evolutionary}  progenitor of \protect\cite{jones_2013}. The results are similar to those of the isentropic star. }
\label{fig:Jones velocity}
\end{center}
\end{figure}

Trajectories of mass elements as a function of time are shown in Fig.  \ref{fig:Jones rtplot}. 
A remnant of $1.23\solarmass$ is left. Approximately $0.14\solarmass$ is ejected, most of it immediately after bounce.
The ejecta eventually expands homologously with velocities of about $2$-$3\times10^{9}{\mathrm{cm}}~{\mathrm{sec}^{-1}}$ (see Fig. \ref{fig:Jones velocity}).
As in the case of the isentropic   progenitor,  a tiny amount of mass travels at much larger velocities.

\begin{figure}\begin{center}\includegraphics[scale=0.57]{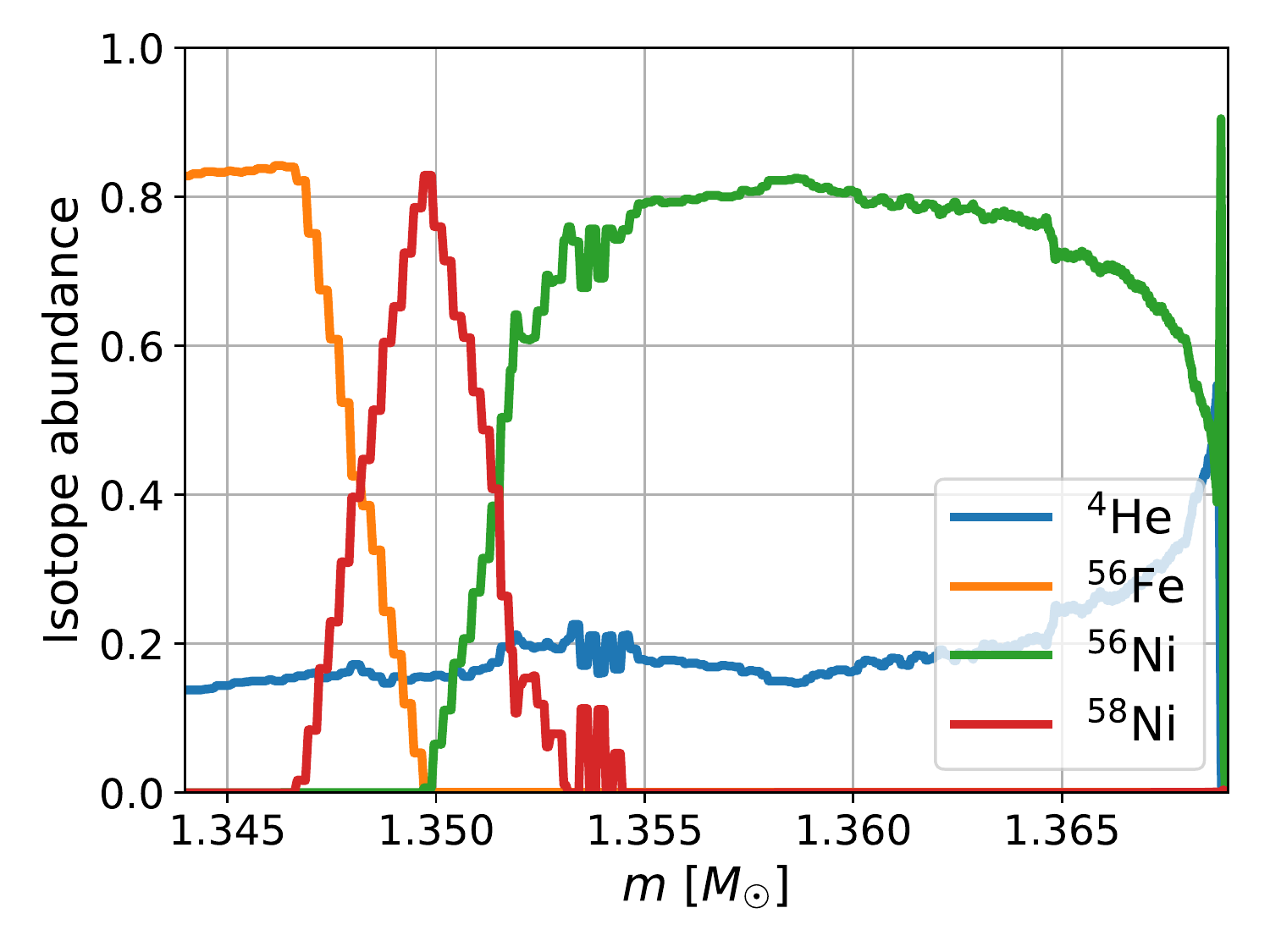}\vskip-0.5cm\caption{The final composition of the ejecta for the evolutionary  progenitor {of~\protect\cite{jones_2013}}. The results are similar to those of the isentropic   star.}\label{fig:Jones composition}\end{center}\end{figure}

The composition of the {outermost region of the} ejected mass at the final time of the simulation, long after it froze out, is shown in Fig. \ref{fig:Jones composition}. As in the isentropic   star  case, we see an outer shell of $\approx 0.02\solarmass$ which is composed mostly of $^{56}$Ni, followed by a small shell of $^{58}$Ni. Further in, there is  a large bulk of $^{56}$Fe, although, as discussed in~\S\ref{subsection:isen ejected mass}, this $^{56}$Fe bulk is in fact composed of various other iron group isotopes. 
As before, the outer shell in which the $^{56}$Ni lies has an electron fraction $Y_e\approx 0.5$, while the main bulk has lower electron fractions $Y_e\approx 0.4$.

\subsection{An evolutionary  binary-star progenitor}
\begin{figure}
\begin{center}
\includegraphics[scale=0.34]{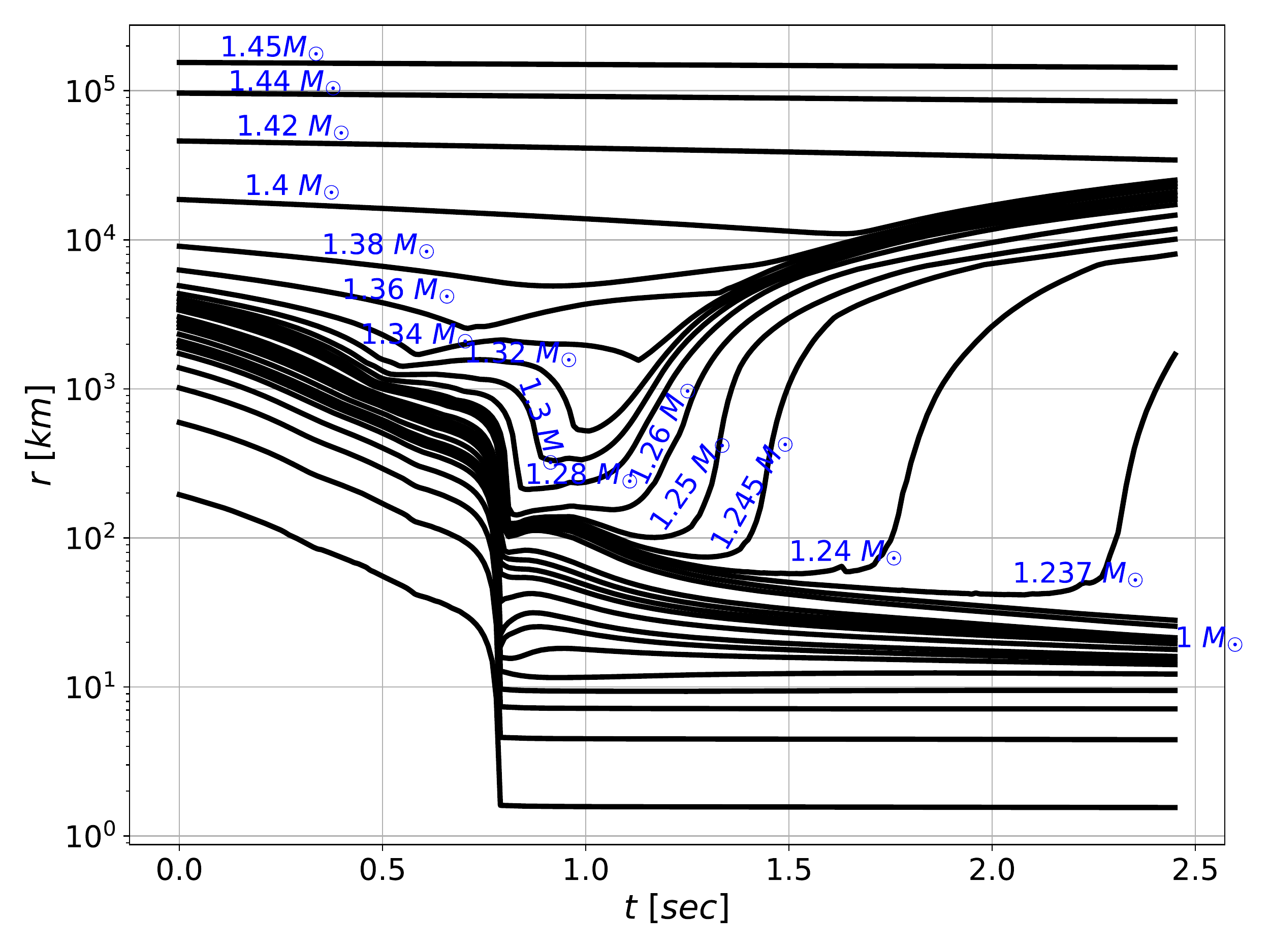}
\caption{Trajectories of  mass elements for the evolutionary  binary-star progenitor of~\protect\cite{tauris_2015}. { The collapse was induced by providing the progenitor with an inwards velocity.} A progenitor of mass $\approx 1.45\solarmass$ collapses, and a PNS of mass $1.24\solarmass$ is formed. The outer layers of the progenitor, as well as the tenuous envelope which did not move significantly during the collapse, are ejected. Shock breakout did not occur during the entire simulation and is expected at time $\approx 10$ sec. The results are similar to other progenitors we studied.}
\label{fig:Tauris rtplot}
\end{center}
\end{figure}

We also simulated the collapse of a  progenitor calculated by \cite{tauris_2015} which was evolved as a binary companion of a NS. 
This progenitor has a slightly larger mass of $\approx 1.45\solarmass$ (including its tenuous envelope). 
{ In their work, \cite{tauris_2015} evolved the progenitor using the BEC code, which is usually suitable to follow the evolution only up to a few tens of years prior to the onset of gravitational collapse~\citep{moriya_2017}. Continuing the evolution, using different codes, shows that the density of the progenitor increases significantly until it collapses (c.f.~\cite{Muller_2018}). } {Consequently}, {  the initial configuration described in~\cite{tauris_2015}} is much larger and   has a significantly lower density compared to the other cases we studied. Therefore, the rate of electron capture in the center of the star was not sufficient to induce its collapse in our simulations{ , when we used this initial configuration}. 
{ Still, using this progenitor is interesting as it will demonstrate that our results are not very sensitive to the initial conditions assumed.}
{ To  calculate the collapse, instead of continuing its evolution we simply} induced the collapse by providing the progenitor with an inwards velocity, resulting in kinetic energy of $\sim 10^{50}$erg. 
We also managed to induce the collapse in a different simulation of this progenitor, by artificially reducing the value of $Y_e$ to $0.2$ for the inner $5\times 10^{-2}\solarmass$ of the progenitor,  at time $0$. The collapse took a little longer to occur in this case (bounce occurred at $t\approx 2.5$s), but the results were very similar. 
Trajectories of mass elements as a function of time are shown in Figure~\ref{fig:Tauris rtplot}.
Due to the large radius of this progenitor, the collapse occurs on a longer time scale and bounce is only at $t\approx 0.7$s.
The outer $\approx 0.05\solarmass$ of the star, which is a tenuous $^{4}\mathrm{He}$-$^{12}\mathrm{C}$-$^{16}\mathrm{O}$ envelope at a radius of  $\sim 10^{5}$km, did not move significantly. Shock breakout did not occur during the entire simulation and is expected at time $\approx 10$sec. Once it occurs this outer region is expected to be ejected as well.

\begin{figure}
\begin{center}
\includegraphics[scale=0.57]{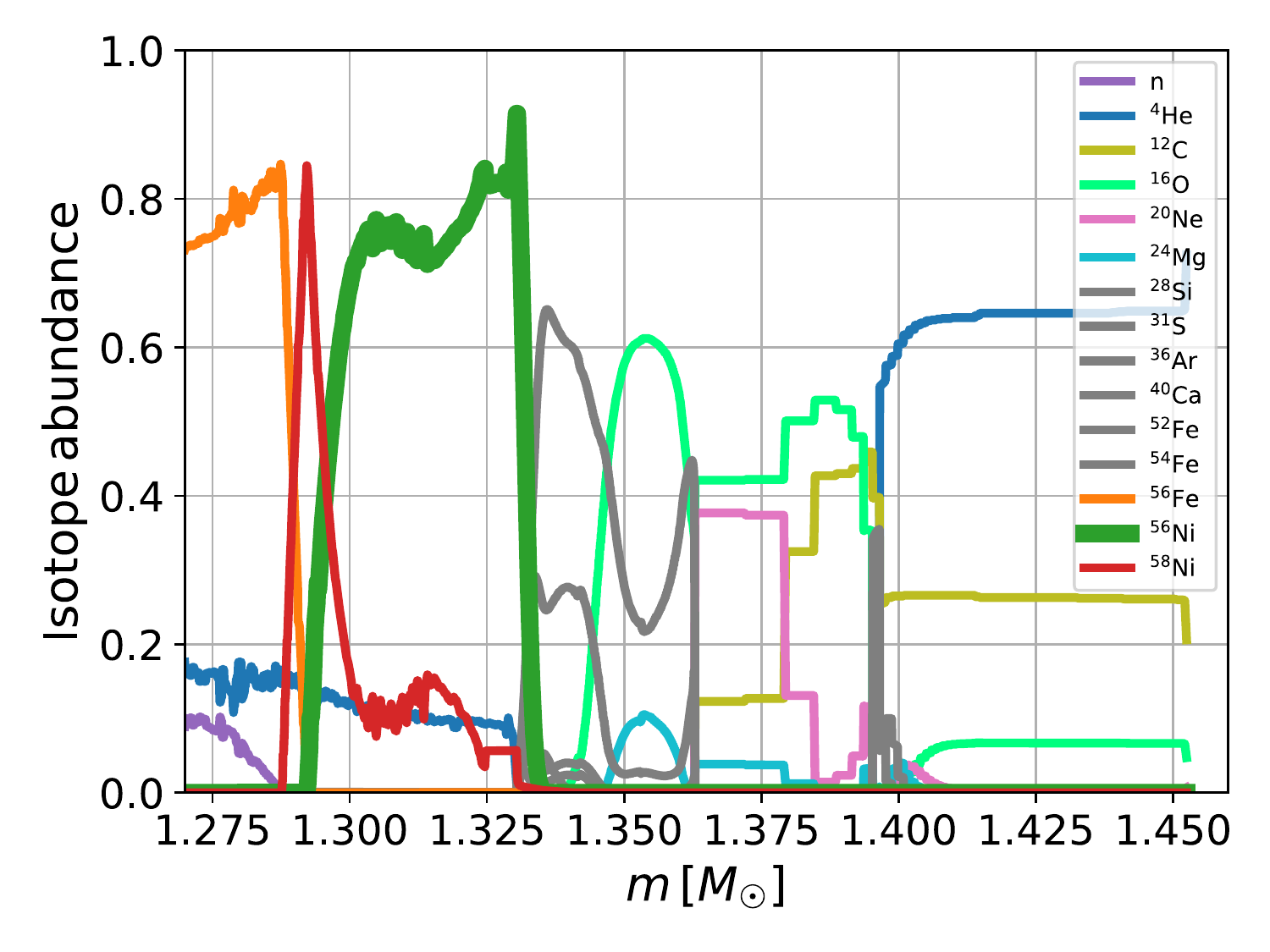}
\caption{
The composition of the ejecta vs the accumulating mass coordinate for the \protect\cite{tauris_2015} progenitor. The divergent shock is located at mass coordinate $m\approx 1.36\solarmass$, and so the outer region is at its initial composition. Below it there is a region which burnt partially and did not reach nickel. In the inner parts the composition is similar to previous cases, with \Ni56 at the outer zones ($\approx 3\times10^{-2}\solarmass$ in this case) and a narrow peak of \Ni58 (see ~\S\ref{subsection:isen ejected mass}).
The \Ni56 abundance is shown in a thick green curve.
}
\label{fig:Tauris composition}
\end{center}
\end{figure}

Still, despite all of these differences, the collapse resembles the cases studies in \S\ref{section:isentropic stars}-\S\ref{section:jones}. A NS of $\approx 1.24\solarmass$ formed. The electron fraction is $Y_e\approx0.5$ for the outer parts of the ejecta, 
and $Y_e\approx0.4$ for the inner parts. 
At the final time of our simulation, the divergent shock is located at mass coordinate $m\approx 1.36\solarmass$. Therefore, as shown in Figure~\ref{fig:Tauris composition}, the composition at the outermost part is the original,  $^{16}$O-$^{20}$Ne-$^{24}$Mg {up to mass $\approx 1.39\solarmass$}, and $^{4}$He-$^{12}$C-$^{16}$O for the envelope. 
Below it there is a region which did go through nuclear burning but did not reach nickel. In the inner parts, the composition is similar to previous cases, with \Ni56 at the outer zones ($\approx 3\times10^{-2}\solarmass$ in this case), a narrow peak of $^{56}$Ni, and then a some other {iron peak} elements where $Y_e\approx 0.4$ (see ~\S\ref{subsection:isen ejected mass}).
The velocities range between $1$-$2\times 10^{9}{\mathrm{cm}}/{\mathrm{sec}}$ {for most of the ejecta}. These are somewhat smaller, yet comparable to  those found in the previous cases.   

We conclude that stars which evolve in binaries may also go through bare collapse and form a NS with similar mass, while ejecting mass with comparable properties, hence inducing a similar observed signal.

\subsection{ Different isentropic   progenitors} \label{section:progenitor_dependence}
We used the isentropic progenitor as a model to study the collapse of different stars, focusing on the effect of varying the progenitors' mass.

We found that lighter progenitors with mass smaller than the Chandrasekhar mass by up to $\approx 0.1\solarmass$ (with central density and temperature as in \S\ref{subsection:our progenitors}) may still collapse, with similar ejecta properties. In particular, this allows for the formation of even lighter NSs.
{ On the other hand, light enough stars { are } stable and do not collapse nor produce NSs.  The particular point of transition from stable to unstable WD (given accurate EOS, reaction rates and neutrino cross sections) is unclear and should be further investigated.}

\begin{figure} 
 \begin{center}
\includegraphics[scale=0.57]{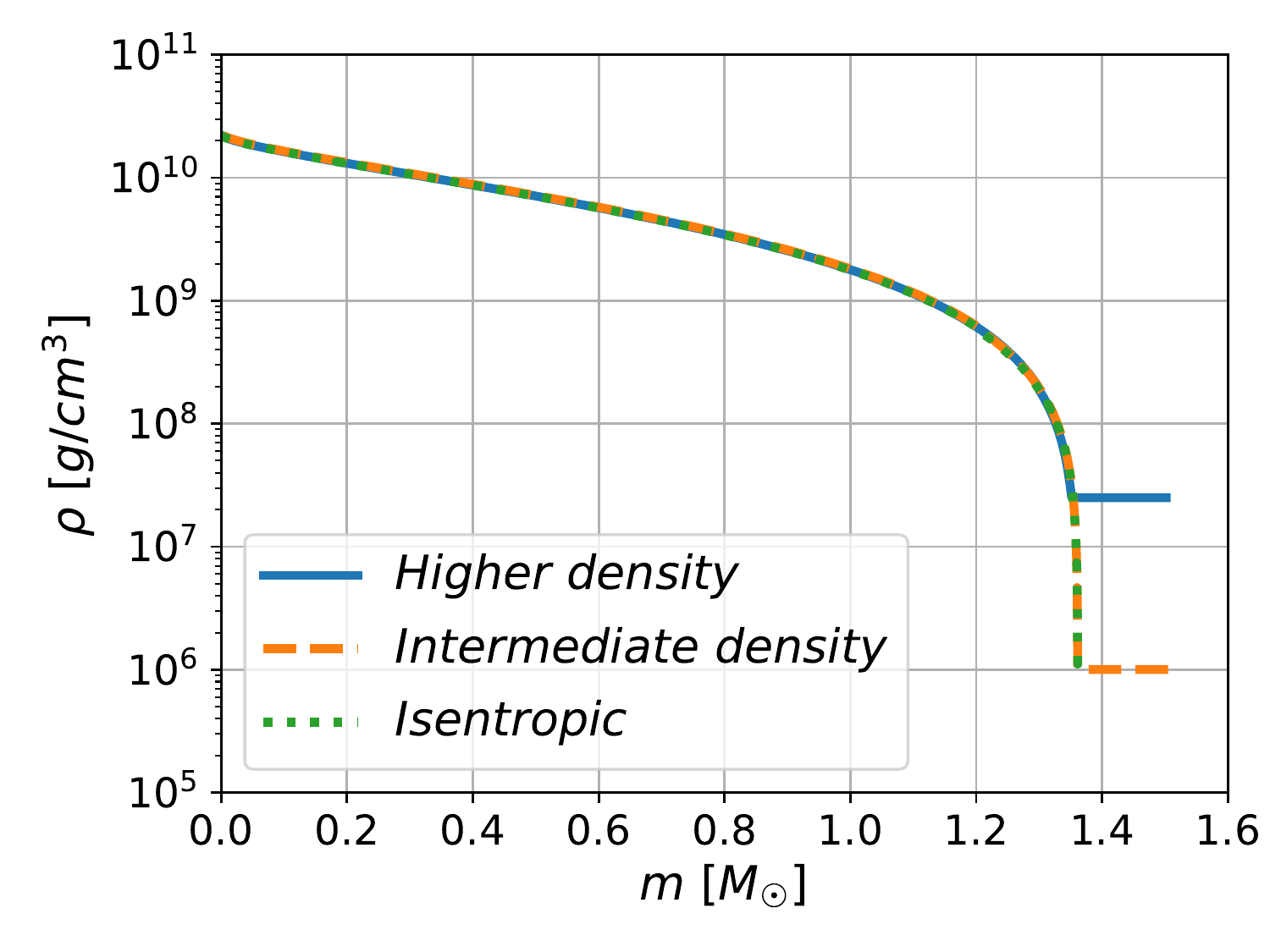}
\vskip-0.3cm
\caption{{ Initial density for two modified $1.5\solarmass$ progenitors. One with a higher density envelope (blue) and another with an intermediate density envelope (orange), compared to the Chandrasekhar mass  isentropic progenitor (green). {The density is the same as the isentropic progenitor until some point,  {$m_0$,} slightly below $ 1.35\solarmass$. 
From this point, we add an outer { uniform} shell whose physical properties are the same as those of the mass element at $m_0$. To allow envelopes of different densities, we simply change $m_0$ by a small amount.}
}}
\label{fig:modified_progenitors}
\end{center}
\end{figure}

Turning to heavier progenitors, we considered progenitors of approximately Chandrasekhar mass, with an outer envelope added on top. 
{ These progenitors are the same as the original Chandrasekhar-mass isentropic  progenitor up to {   some point which is approximately at}  $m_0\approx 1.35\solarmass$, { namely slightly below the original boundary}. From this point, we add an outer { uniform} shell whose physical properties are the same as those of the mass element at $m_0$. To allow envelopes of different densities, we simply change $m_0$ by a small amount.
{ Some examples of initial configurations of heavier progenitors are shown in Figure~\ref{fig:modified_progenitors}.}}

{ 
We note that adding the mass in this way can have significant and non trivial impact on the fate of the progenitor.  For example instead of  going through ECSN it may explode as a type Ia SN. We don't consider these possibilities here. Instead, we only try to shed some light on the possible consequences of collapse of heavier progenitors  assuming ECSN occurs in these initial conditions, without discussing the extent to which these conditions are valid to describe realistic more massive ECSN progenitors. As such these results should be taken with a grain of salt and clearly the question what is the structure of heavier progenitors should be explored more extensively. }
{ With all that being said, one possible realistic scenario for ECSNe of WDs with envelopes could rise from the merger of two WDs, as suggested by~\cite{Lyutikov_2019,Lyutikov_2022} (others may include the partial shedding of a heavier envelope in the pre-collapse evolution due two stellar wind or interaction with a binary). So} although  these simplified models are artificial from a stellar evolution point of view, they can give us indications on what happens in realistic configurations of this nature. 

\begin{figure} 
\begin{center}
\includegraphics[scale=0.34]{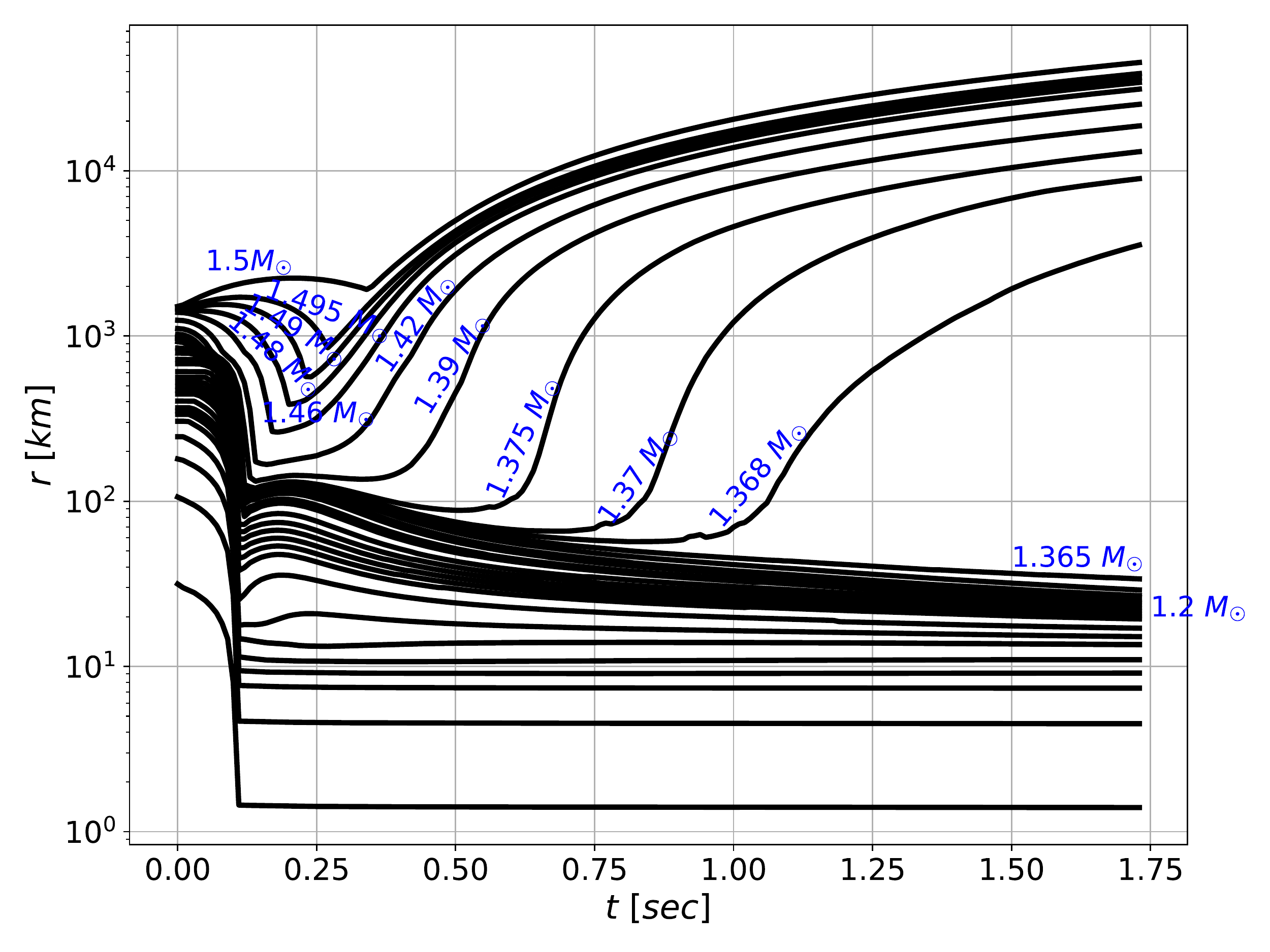}
\caption{{ Trajectories of mass elements for a progenitor with a high density envelope, and total mass $\approx 1.5\solarmass$. The progenitor undergoes collapse similar to previous cases, with $\approx 0.14\solarmass$ ejected. This results in the formation of a $1.37\solarmass$ NS, { which is } heavier compared to all cases studied in previous sections. There is a slight early expansion of the very outer mass element near the boundary of the progenitor, caused by the initial progenitor being out of hydrostatic equilibrium. }}
\label{fig:heavy NS rtplot}
\end{center}
\end{figure}

\begin{figure}\begin{center}\includegraphics[scale=0.57]{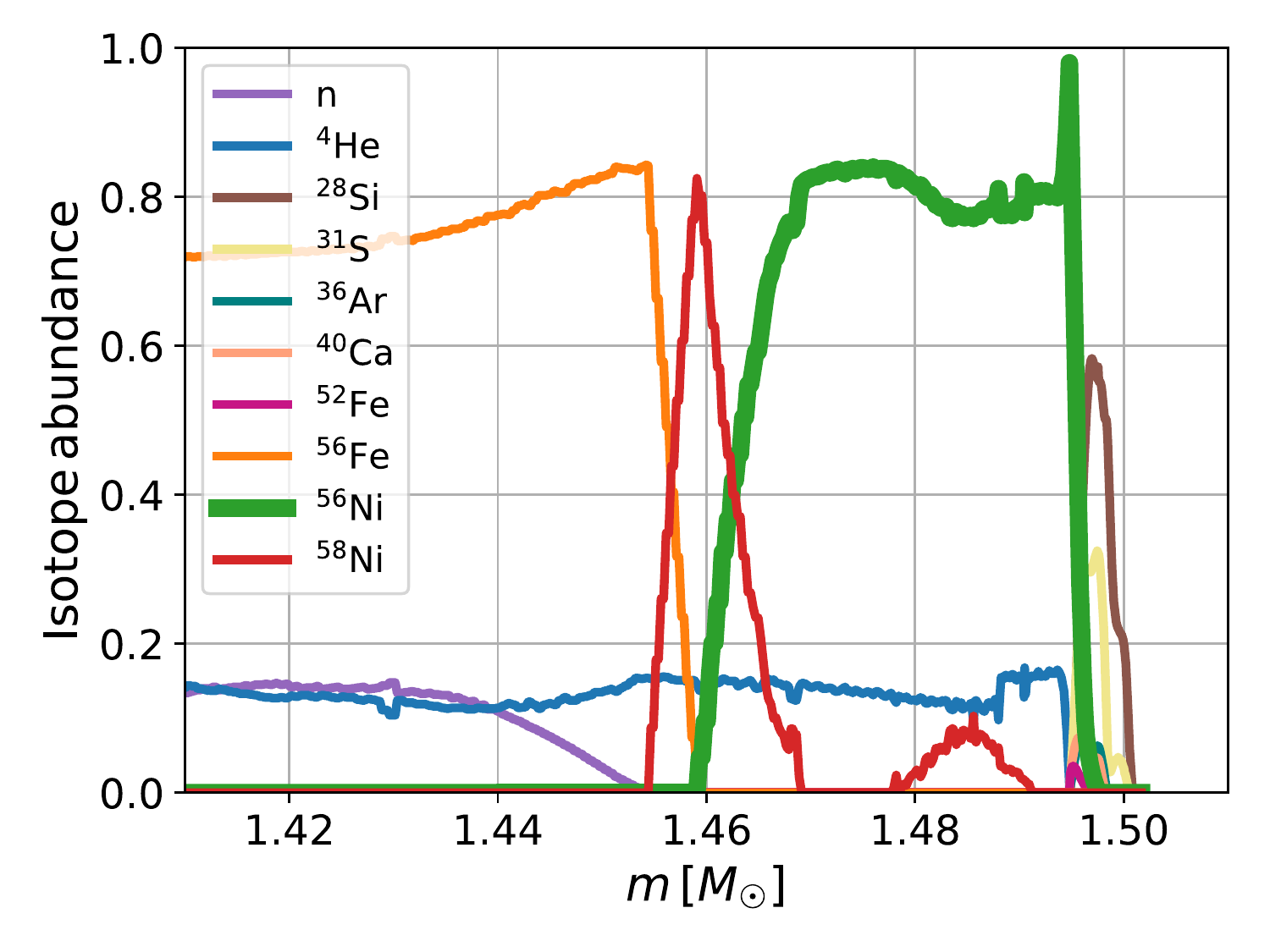}\vskip-0.5cm
\caption{{ The final composition of the ejecta for a progenitor with a high density envelope, and total mass $\approx 1.5\solarmass$. The \Ni56 mass, $\approx 3\times 10^{-2}\solarmass$,  and the composition are similar to { those found in } previous studied cases.}}
\label{fig:heavy NS composition}
\end{center}
\end{figure}

\begin{figure} 
\begin{center}
\includegraphics[scale=0.57]{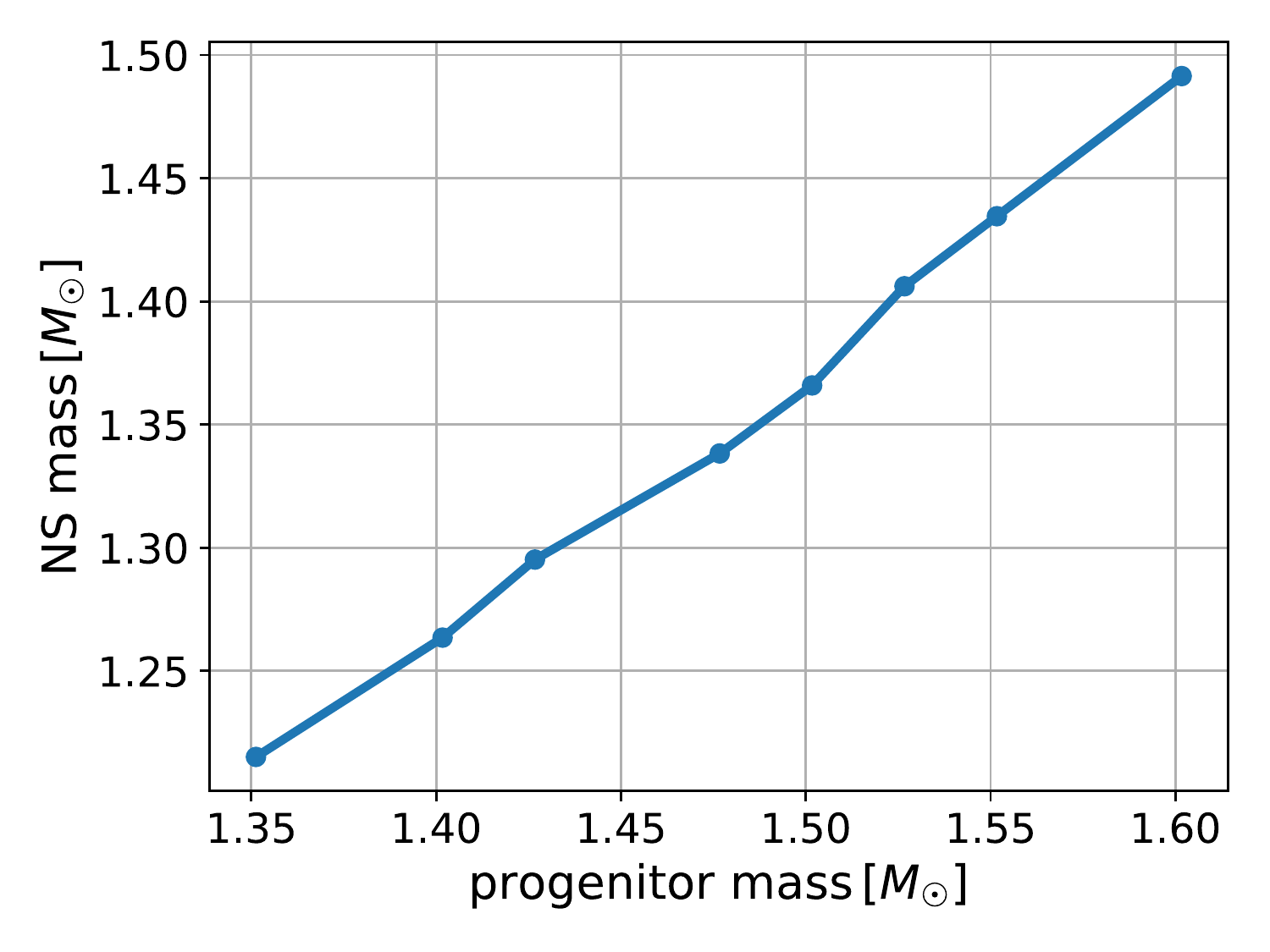}
\vskip-0.3cm
\caption{The remnant NS mass vs the progenitor mass. It is approximately linear with slope $1$, meaning that excess mass in the progenitor stays bound as part of the NS. This occurs for progenitors with dense enough envelopes.}
\label{fig:m_NS vs m_progenitor}
\end{center}
\end{figure}

For progenitors with envelopes whose density is above a certain value, our simulations resulted in a collapse leading to similar ejecta properties, mass and composition,  as in~\S\ref{section:isentropic stars}, independent of the mass. Specifically, for the particular case of $\rho_{env}\approx 10^{7}\frac{\mathrm{g}}{\mathrm{cm}^3}$, and envelope mass up to $0.25\solarmass$, we found mass ejection of $0.11$-$0.14\solarmass$, of which $2$-$3\times10^{-2}\solarmass$ was composed of \Ni56, at ejecta velocities of $1$-$3\times10^{9}{\mathrm{cm}}/{\mathrm{sec}}$. { Trajectories of mass elements as a function of time, and the composition of the ejecta, for a $\approx1.5\solarmass$ progenitor, are shown in Fig.~\ref{fig:heavy NS rtplot} and Fig~\ref{fig:heavy NS composition} respectively.} The resulting NS is, of course, correspondingly heavier. Fig.  \ref{fig:m_NS vs m_progenitor} shows that the NS mass increases linearly with the  progenitor mass with slope approximately $1$.
As the mean mass for NSs in BNS systems is $\approx 1.33\solarmass$ \citep{ozel_2016}, these calculations show that bare collapse is consistent with the formation of BNS systems in this regard as well.


In the other limit, where the  envelope  density is very small (in particular lying at large radii), the envelope does not affect the collapse or the resulting NS significantly. Instead, the envelope is simply ejected, without undergoing nuclear burning.

\begin{figure} 
\begin{center}
\includegraphics[scale=0.34]{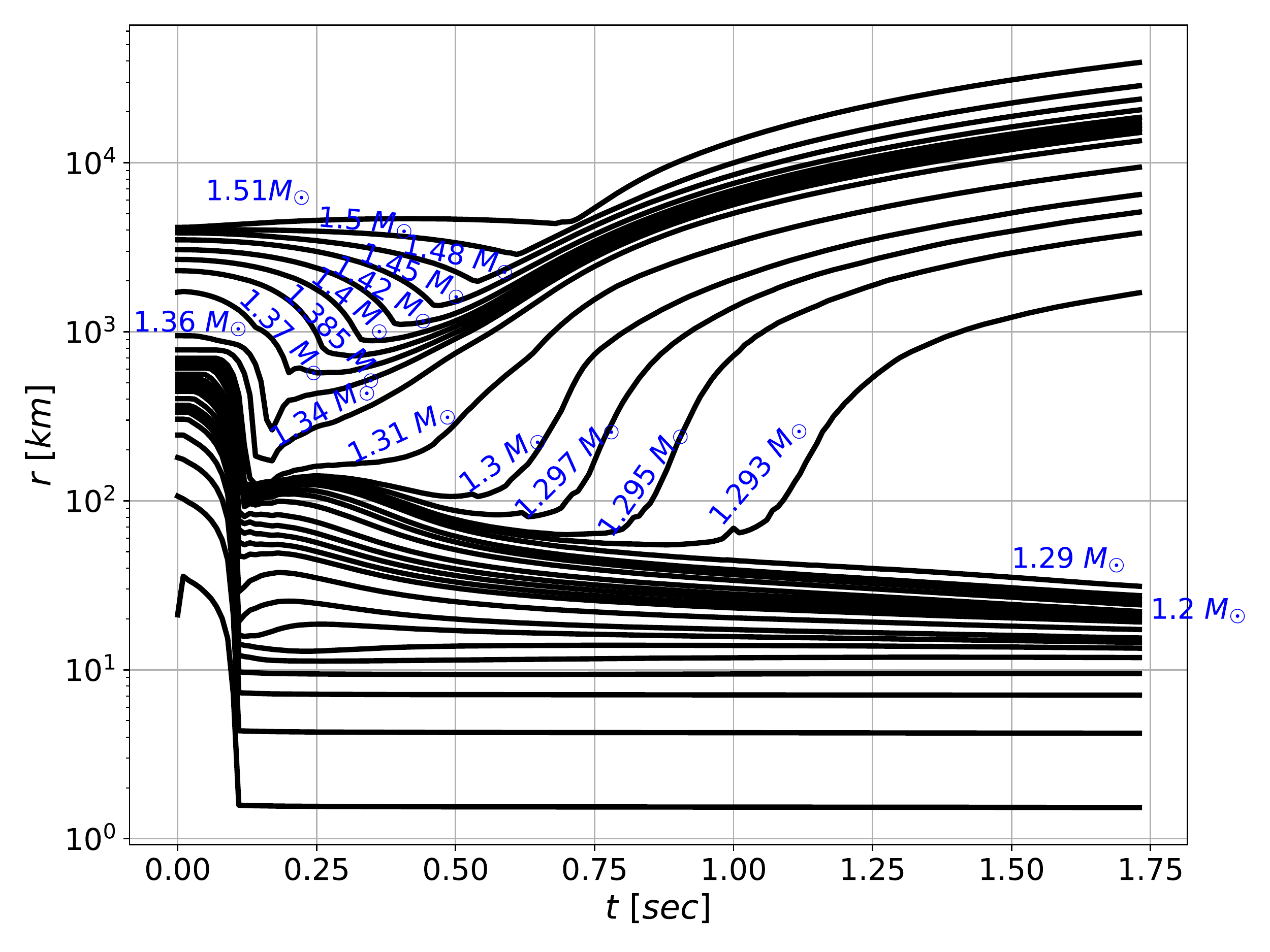}
\caption{{ Trajectories of mass elements for a progenitor with an intermediate density envelope, and total mass $\approx 1.5\solarmass$. The progenitor undergoes collapse similar to previous cases and result in the formation of a $1.29\solarmass$ NS. Most of the additional envelope mass is ejected, yet some of it remains bound and the NS is slightly heavier compared to the cases studied in previous sections. In this intermediate density  envelope density case, the \Ni56 mass is significantly larger.}}
\label{fig:high Ni rtplot}
\end{center}
\end{figure}

\begin{figure}\begin{center}\includegraphics[scale=0.57]{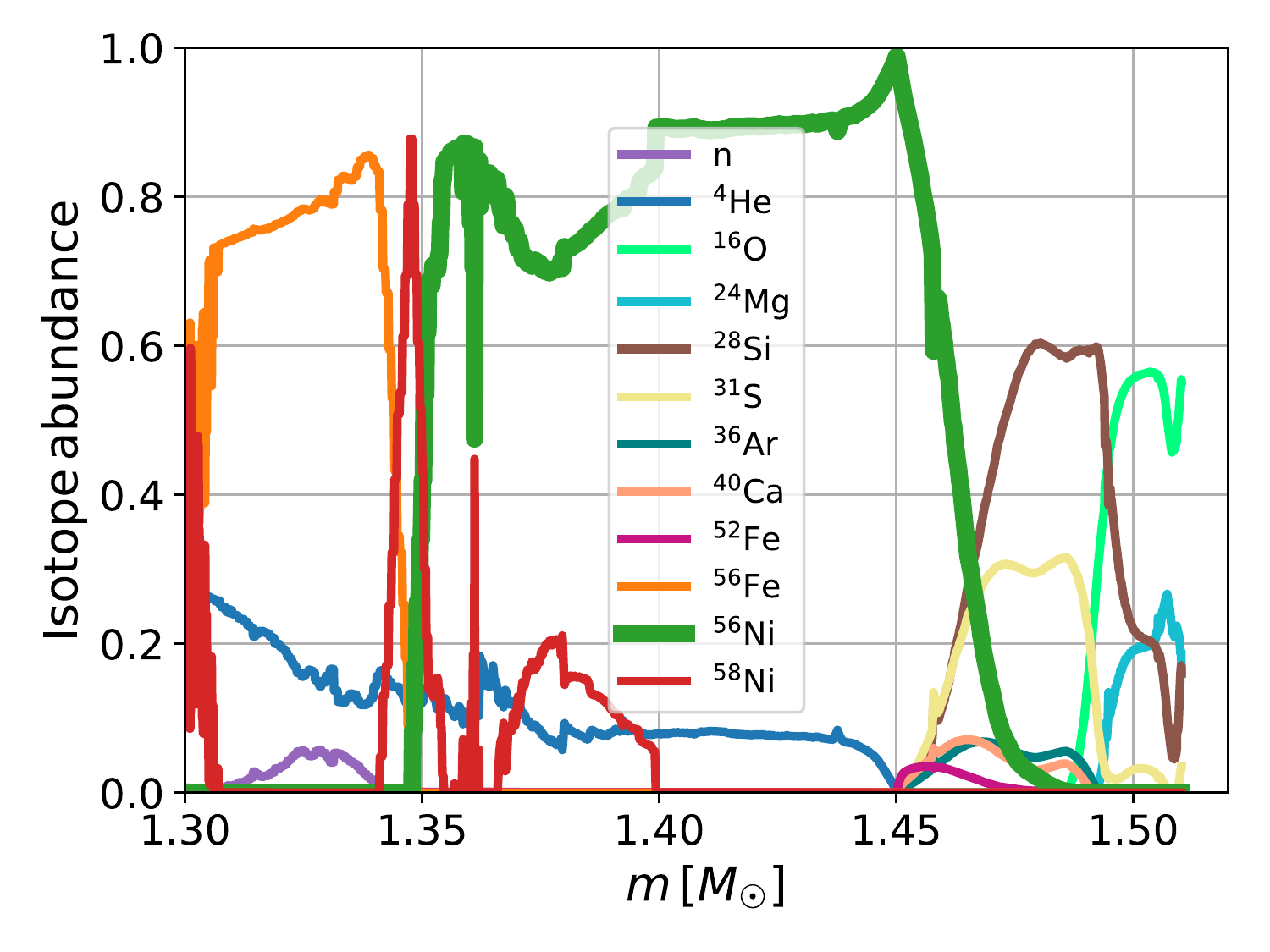}\vskip-0.5cm
\caption{{ The final composition of the ejecta for a progenitor with an intermediate density envelope, and a total mass $\approx 1.5\solarmass$. The \Ni56 mass is $\approx 9\times 10^{-2}\solarmass$, significantly larger than previously studied cases. The nucleosynthesis results were also validated using {\sc SkyNet}.}}
\label{fig:high Ni composition}
\end{center}
\end{figure}

\begin{figure}
\begin{center}
\includegraphics[scale=0.57]{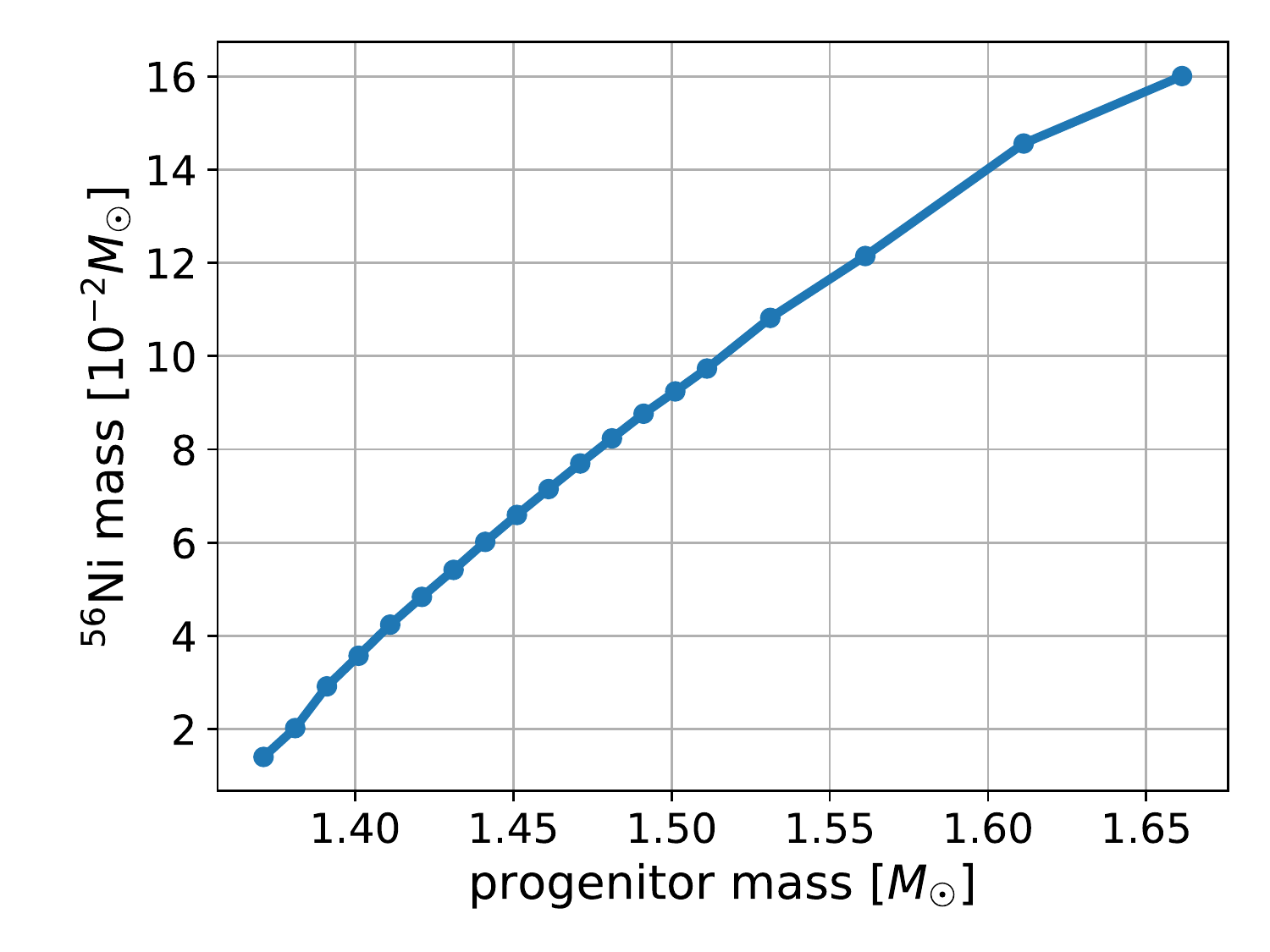}
\vskip-0.5cm
\caption{The $^{56}$Ni mass in the ejecta vs the progenitor mass. About  $1/2$ of the added mass  is ejected as $^{56}$Ni. The nucleosynthesis results were also validated using {\sc SkyNet}. {This occurs for progenitors with envelopes of a limited range of densities.}}
\label{fig:increase_Ni m_Ni_vs_m_prof}
\end{center}
\end{figure}

In the intermediate  density regime, we were able to find cases where the envelope was on the one hand dense enough to collapse (and in particular reach the suitable thermodynamic state for nuclear burning), but on the other hand is light enough for most of the mass that was added to eventually be ejected upon the bounce.
Figure~\ref{fig:increase_Ni m_Ni_vs_m_prof} depicts the $^{56}$Ni mass as a function of the progenitor mass, for the particular case of $\rho_{env}\approx 10^{6}{\mathrm{g}}/{\mathrm{cm}^3}$. It can be seen that about {$1/2$} of the added mass is converted to \Ni56. The largest progenitor we considered, with a mass of {$1.66\solarmass$}, produces {$16\times 10^{-2}\solarmass$} of \Ni56,  {$5$}-{$8$} times larger than the amount formed in the Chandrasekhar-mass progenitors. 
We also validated these nucleosynthesis results using {\sc SkyNet}, by the method discussed in~\S\ref{subsection:isen ejected mass}, and found an excellent agreement.
For these progenitors, the mass of the formed PNS was $1.25$-$1.3\solarmass$. { Trajectories of mass elements as a function of time, and the composition of the ejecta, for a $\approx 1.5\solarmass$ progenitor, are shown in Fig.~\ref{fig:high Ni rtplot} and Fig~\ref{fig:high Ni composition} respectively.}
{  The amount of \Ni56 formed in this particular intermediate density regime progenitors{ , $ 9 \times 10^{-2}\solarmass$,} is considerably larger from all previous simulations of ECSNe. For instance, \cite{sawada_2021} argued that is difficult to achieve \Ni56 mass larger than $5\times10^{-2}\solarmass$. 
The  particularly high amount of \Ni56 in our simulations may arise from  the specific envelope density we have chosen (together with the simplified model of uniform envelope density). 
While such models may not describe correctly  realistic  ECSN progenitors, they still  suggest that some unique progenitors could synthesize more \Ni56 than previously estimated.}


\section{ Observational Signature}
\label{sec:observations}

\subsection{Fast Transients}
\label{sec:Fastransients}
 The gravitational collapse of bare degenerate cores  results in many cases in the ejection of  $\sim 0.1\solarmass$,  at about $0.1c$. The outermost layer of this ejecta is composed of $\sim 0.02\solarmass$ of $^{56}$Ni.   
The decay of this \Ni56 will produce an ultra-stripped SN with very different characteristic luminosity and rise time fro a typical SN. 
We turn now to  consider the observational characteristics of such an outer thin shell of radioactive nickel, as it expands homologously.
We use a simple model that gives an order of magnitude estimate of the observed bolometric luminosity, peak time and temperatures. 

Consider a shell of mass $M$, radius $R$ and width $\Delta R$, that is moving homologously so that $R=v t $ and $\Delta R\propto t$.
Let $M_{^{56}\mathrm{Ni}}$ be the mass  of $^{56}$Ni  within the shell and $\kappa$ the opacity of the shell.
The optical depth across the shell is given by
\begin{equation}
\tau= \int_{R-\Delta R}^{R} \kappa\rho dr\approx
\frac{\kappa M }{ 4 \pi (v t)^2 } \ ,
\label{eq:tau}
\end{equation}
where we have assumed that the opacity is constant, the density does not depend too strongly on $r$, and  $\Delta R\ll R$. Since the \Ni56 is arranged in a thin shell this  reduces $\tau$ by a factor of $3$, compared to a sphere.

Radiation  escapes at time $t$  only from regions whose optical depth $\tau^{\prime}$ satisfies  ${c}/{\tau^{\prime}}> v$.  At early times, the  effective observable mass is $M_{\mathrm{obs}}(t)={4\pi c v t^2}/{\kappa}$ and the bolometric light curve rises like $ t^{2}$. The luminosity peaks when radiation can escape from the whole shell, namely at:
\begin{equation}
t_{\mathrm{peak}}= \sqrt{ \frac{\kappa M }{4 \pi c v}} \approx  0.5 ~{\rm days}~ \frac{\kappa_{-1}^{1/2} M_{-2}^{1/2}}{v_{-1}^{1/2}} \ ,
\label{eq:t_rise}
\end{equation}
where we use the notation $Q_x=Q/10^x$ in cgs units but masses are measured in $\solarmass$, velocities in $c$ and time in days. 

For $t_{\mathrm{peak}}\ll \tau_{^{56}\mathrm{Ni}}$ 
the peak luminosity is:
\begin{equation}\label{eq:L_peak}
L_{\mathrm{peak}}\approx \frac{ M_{^{56}\mathrm{Ni}} \epsilon_{^{56}\mathrm{Ni}}} {\tau_{^{56}\mathrm{Ni}}} \approx  7.7\times 10^{41}\times M_{^{56}\mathrm{Ni},-2}{\mathrm{erg}}/{\mathrm{s}} \ ,
\end{equation}
where $\epsilon_{^{56}\mathrm{Ni}}\approx 2.96\times 10^{16}$ erg/g is the energy released per gram of decaying \Ni56, $\tau_{^{56}\mathrm{Ni}}\approx 8.8$ day is the mean lifetime of \Ni56, and $M_{^{56}\mathrm{Ni},-2}=M_{^{56}\mathrm{Ni}}/(10^{-2}\solarmass)$.
The effective temperature, 
\begin{equation}
\label{eq:peakTemperature}
    T_{\mathrm{peak}}=\Big(\frac{L_{\mathrm{peak}}}{4\pi (vt_{\mathrm{peak}})^2 \sigma}\Big)^{1/4}\approx 1.6\times 10^{4}~ \kappa_{-1}^{-1/4} v_{-1}^{-1/4} \Big(\frac{M_{^{56}\mathrm{Ni}}}{M}\Big)^{1/4}\  K \ , \  
\end{equation}
corresponds to a UV/blue signal. The spectrum will show lines according to the specific composition typical for bare collapse, which is mostly iron group and lighter elements. We expect that 
 different bare collapses  should  have  comparable spectra.

After the peak, the luminosity decreases due to a combination of the decrease in the amount of the decaying \Ni56, and the leakage of the  decay products out from the expanding system. The latter is more important, leading to a significantly faster decline than the one arising from  the \Ni56 lifetime. We expect  the whole system to be optically thin long before we can see the typical radioactive decay time of \Ni56.

A typical value of $M_{^{56}\mathrm{Ni}} =0.02$-$0.03\solarmass$ of $^{56}$Ni is sufficient to explain a fast transient with a peak luminosity of $\approx 2\times 10^{42}{\mathrm{erg}}/ {\mathrm{s}}$ or equivalently a peak absolute magnitude of $\approx -17$.  The largest values of \Ni56\ that we have found  of $\approx 0.16\solarmass$  yields a peak absolute magnitude of $\mathcal{M}_{\mathrm{peak}}\approx -19$. The typical rise time of these events is  1-2 days.

\begin{figure}
\begin{center}
\includegraphics[scale=0.65]{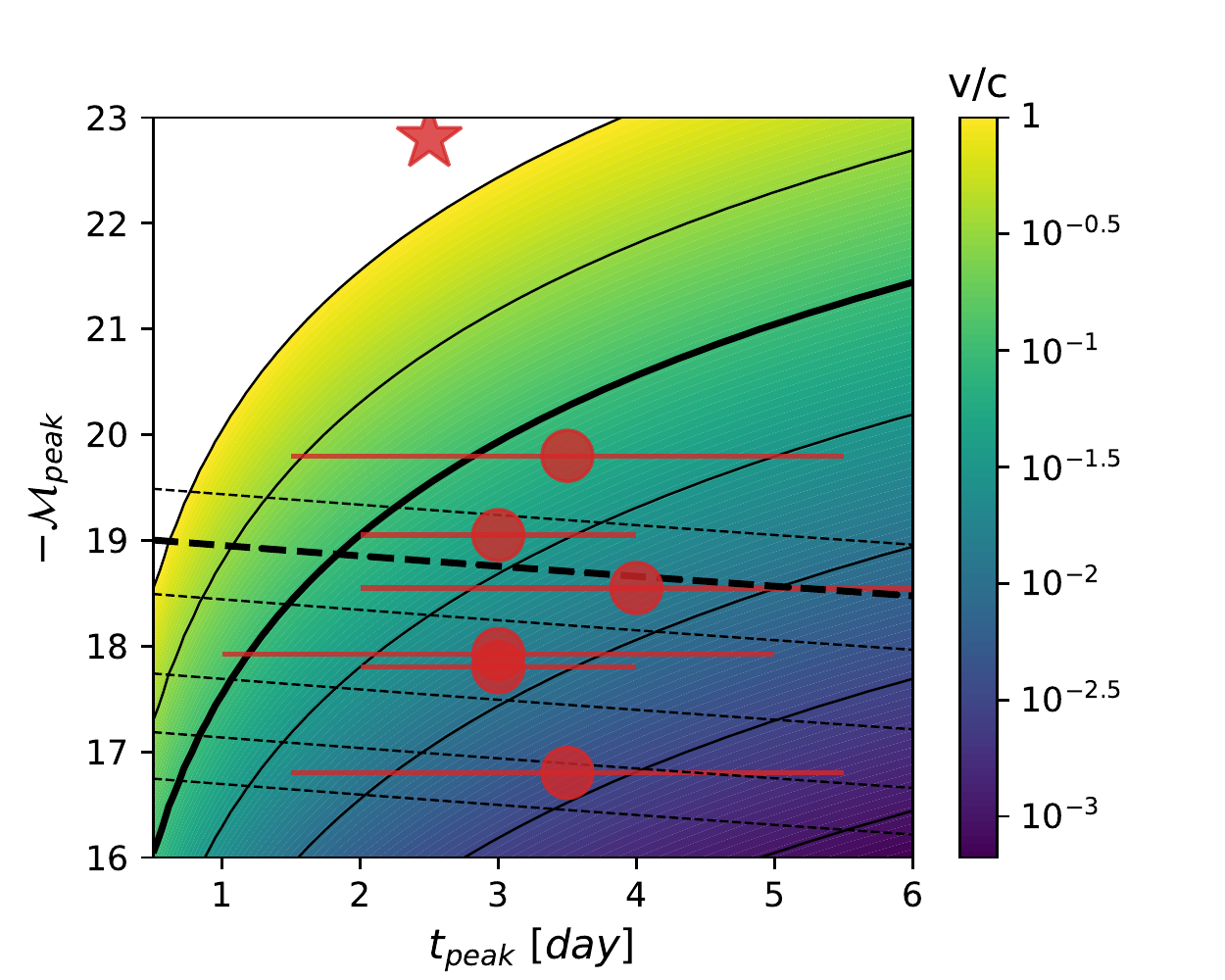}
\caption{Contour map of the minimal required velocity for given peak time and peak absolute magnitude{ , obtained { for  a given  outer shell of $M_{^{56}\mathrm{Ni}}$, namely $M=M_{^{56}\mathrm{Ni}}$ and  $\kappa_{-1}=1$}}. The black curves (full lines) correspond to the values\linebreak $v/c=0.001,0.003,0.01,0.03,0.1,0.3,1$. Each such curve defines the most luminous events theoretically possible by our model, for a given velocity $v$. The black dashed lines show the value of $\mathcal{M}_{\mathrm{peak}}$ for {$M_{^{56}\mathrm{Ni}}=0.02,0.03,0.05,0.1,0.16,0.25 \solarmass$}.
The bold full curve corresponds to $v/c=0.1$ which is the typical velocity for most of the ejecta, and the bold dashed line corresponds to \mbox{$M_{^{56}\mathrm{Ni}}=0.16 \solarmass$} which is the highest $^{56}$Ni mass obtained in our simulations. {Also shown in red are some of the PS1 fast transients~\citep[][ as circles]{drout_2014}  and AT 2018cow (as a star).}
}
\label{fig:v}
\end{center}
\end{figure}

Combining  Eqs.~\eqref{eq:t_rise} and  \eqref{eq:L_peak} we obtain the ejecta velocity needed for a given rise time, and  peak absolute magnitude:
\begin{equation}
    v_{-1}=0.25\kappa_{-1}\frac{M_{-2}}{t_{\mathrm{peak},1}^2}\approx 3.2k_{-1}\frac{M}{M_{^{56}\mathrm{Ni}}}\frac{L_{\mathrm{peak},43}}{t_{\mathrm{peak},1}^2} \ . 
\end{equation}
The minimal required velocity to obtain a transient of given peak time $t_{\mathrm{peak}}$ and peak absolute magnitude $\mathcal{M}_{\mathrm{peak}}$  for a given  $M_{^{56}\mathrm{Ni}}$ is shown in Fig.  \ref{fig:v}{ , assuming $\kappa_{-1}=1$}. This minimal velocity is obtained when $M=M_{^{56}\mathrm{Ni}}$. Also shown is the  dependence of $\mathcal{M}_{\mathrm{peak}}$ on $t_{\mathrm{peak}}$ for different $^{56}$Ni masses. Together, this outlines the region in which fast flares can be explained by bare collapses. 
These values can explain many of the fast transients and in particular most of   the ``golden sample" of  {\cite{drout_2014}}.

\subsection{Other possible observational signatures}
\label{sec:othertransients}

\subsubsection{Very luminous fast transients}

{The luminosity arising from \Ni56 decay in bare collapses can explain many of the fast transients. 
For example it is sufficient to explain the ``golden sample" discussed by  \cite{drout_2014}. 
However, it falls short of the brightest ones like  AT 2018cow whose peak luminosity is $\sim 4 \times 10^{44}$erg/s \citep{Perley2019} or  the brightest fast transients reported by {\citep{drout_2014}}  that would require  more than $0.3\solarmass$ of \Ni56.}

The ejecta of  $\sim 0.1\solarmass$ travelling at   $\sim 0.1 c$, has a significant kinetic energy of the order of   $\sim 10^{51}$ erg. This  is comparable to the kinetic energy of a typical (much more massive) SN. 
Tapping even a small fraction of this energy can lead to signal that  explains the brightest transients. This could happen if on a time scale  of a day or so,  
the ejecta collide with material in the proximity of up to $\approx 0.1$ light day from to the progenitor. Such material could be  mass  ejected from the progenitor just prior to the collapse. While we cannot demonstrate that such mass ejection takes place on this time scale, given the essential extensive mass loss of  progenitors for bare collapse  and the fast evolution during the last stages before the collapse,   this is clearly possible. 
Assuming the material with which the ejecta collides had a velocity $v_{\mathrm{w}}$, it would have needed to escape the progenitor up to $\approx 10~[v_\mathrm{w} /({3000\mathrm{ km/sec)}}]^{-1}$ days prior to  the collapse.

If the mass of the ejecta,  $\sim 0.1 \solarmass$, is smaller than  the mass with which it collides, this collision could transfer a large fraction of  the kinetic energy to thermal  energy. This is about $100$-$1000$ times   larger than the total energy released by the radioactive decay of the $^{56}$Ni. This could easily power  even the very rare brightest events.
It is important to note that much smaller mass is needed for this scenario to happen, as compared to a similar situation with standard SNe which eject much more mass. Moreover, this scenario can fit nicely  with the evolution of bare collapse progenitors which lose most of their envelope.

\subsubsection{The Remnants}

In the long run, bare collapse events produce a SN remnant (SNR) like signature.  As the ejecta expands it sweeps up mass. Initially it is  coasting at a constant velocity and as it accumulates more and more mass the luminosity increases like $t^2$. The Sedov-Taylor~(ST) phase begins at
\begin{equation}
t_{\mathrm{ST}} =   \big(\frac{ 3 M_{\mathrm{ej}}} {4 \pi \rho_{\mathrm{ISM}} v_{\mathrm{ej}}^3} \big)^{1/3} \approx 40~ {\rm years}~~ {M_{\mathrm{ej,-1}}^{1/3} {n_{\mathrm{ISM}}^{1/3} {v_{-1}} }} \ ,\end{equation}
where $n_{\mathrm{ISM}}$ is the interstellar medium (ISM) density. 
At this stage  the ejecta have passed through a mass of ISM equal to $M_{\mathrm{ej}}$. 
The energy dissipation rate is maximal at  $t_{\mathrm{ST}}$.
As the mass of the ejecta is much lower than the mass of a SN ejecta  and its velocity is  faster, $t_{\mathrm{ST}}$ is shorter by about a factor of $\approx 10$, than in a regular SN. This implies that 
the peak bolometric luminosity of this bare-collapse SNR, $\approx 10^{42}$ erg/sec, is  larger than the  peak luminosity of a regular SNR by a similar factor. This luminosity  is comparable to the peak luminosity of a rare SNR arising from a superluminous SN.  Over even  longer time scale the Sedov-Taylor phase continues and it will be impossible to distinguish this remnant from  a regular SNR.  In this  phase all that matter is the total energy and the external density and the initial ejecta mass and velocity are not relevant. 

\section{Discussion}
\label{sec:discussion}
{We have shown that the collapse of a bare stellar core of around Chandrasekhar mass, results in a light, $\approx 1.2$-$1.35 \solarmass$, neutron star  and ejection of $\approx 0.1$-$0.2\solarmass$ at $\approx 0.1$c. 
We considered two evolutionary  configurations that were derived from  stellar evolution simulations by \cite{jones_2013} and by \cite{tauris_2015}. Similar results were obtained for an isentropic   initial configuration of hot white dwarf, and even when a light envelope of  $\le 0.35 \solarmass$ was added to it. }

Naturally this neutron star is rather light. This is consistent with the observation that the mean mass of the  neutron stars in BNS systems  is $\approx 1.33\solarmass$ \citep{ozel_2016}. 
With such a small mass ejected, a binary composed of this progenitor and a companion neutron star would remain bound with small eccentricity and small proper motion. The double pulsar PSR J0737-3039 is a prototype of a system that formed in this way~\citep{piran_shaviv_2005}. 
This is also consistent with the observations that BNS systems with lower eccentricity and lower CM motion, that are expected to form in this channel,  involve lighter neutron stars \citep{vdh_2007,vdh_2011}.  

The collapse ejects $\sim 0.1 \solarmass$ of iron group elements with an outer shell composed mostly of \Ni56. This outermost region where \Ni56 was formed had $Y_e=0.5$, consistently with the equal number of protons and neutrons in this isotope. 
This radioactive ejecta results in fast transients for two reasons. First,   the  ejecta velocities are  $\approx 2$-$3\times 10^{9}{\mathrm{cm}}~{\mathrm{sec}^{-1}}$. As the rise time of the light curve is proportional to ${v_{\mathrm{ej}}}^{-1/2}$, this leads to faster transients by a factor of $\approx\sqrt{3}$, 
as compared to traditional SNe with $v_{\mathrm{ej}}=10^{9}{\mathrm{cm}}~{\mathrm{sec}^{-1}}$  \citep[see e.g.][]{arcavi_2016}.
Second, 
the $^{56}$Ni lays in a shell at the outermost  layer of the ejecta. The rise time determined by the expansion of a shell, rather than a  sphere, is faster by a factor of up to $\sqrt{3}$.   
Moreover, the rise time was found to depend on the $^{56}$Ni mass, rather than the entire ejecta mass.
Together these two factors combine to shorten the transients by a factor of $\approx 3$, for a given $^{56}$Ni mass. 
For the particular ejecta we obtained in our simulations, we found rise times  of $0.5$-$2$ days.

Chandrasekhar mass progenitors, both isentropic   and evolutionary  (i.e. evolved in detailed stellar evolution schemes), induce an ejection of  $0.02-0.03\solarmass$ of $^{56}$Ni, resulting in a peak luminosity of  $\approx 2\times 10^{42}{\mathrm{erg}}/{\mathrm{sec}}$ or equivalently a peak absolute magnitude of $\approx -17$. This can explain the lower end of the fast flares.
Even higher  amounts of $^{56}$Ni ejecta were found  in (synthetic)  progenitors that included an additional envelope. The largest  amount of $^{56}$Ni that we have found was  $0.16\solarmass$, corresponding to  a peak absolute magnitude of $-19$, which already explains a large fraction  of the observed fast  transients \citep{drout_2014}. 

If the ejecta collides with a massive shell that was ejected within a few days prior to the collapse then the large kinetic energy of the ejecta, which is of order $10^{51}$erg, can be tapped \citep[see e.g][]{Moriya2013,Blinnikov2017,De2018S,Fox2019,Leung2021}.  This can explain the brighter end of the transients.  In fact the combination of the radioactive signal and  interaction of the fast ejecta with earlier outflow can explain double peaked fast transients, like  iPTF 14gqr-SN 2014ft \citep{De2018S}.

Eventually the ejecta will interact with the surrounding ISM and produce a SNR. 
The SNR will peaks on  a time scale of $\sim 40 $ years  with a  peak luminosity larger than a typical peak luminosity of a regular SNR by a factor of almost ten. However, these are rare events and it will be difficult to catch on in such a phase. The longer term (Sedov phase and later) of this SNR will be similar to a regular SNR.

\section{Conclusions}
\label{sec:conclusions}
Analysis of the properties (eccentricity and proper motion) of Galactic binary pulsars has shown that most neutron stars in these systems formed with  minimal mass ejection and without a kick velocity \citep{beniamini_2016}. 
Motivated by these observations we have explored here bare collapses that arise when a stellar core that has lost all its envelope undergoes electron capture and collapse.  
Our calculation begin with stellar progenitors (evolutionary and synthetic) at the end of their life, we calculate the collapse that arises due  electron capture and following the bounce and mass ejection. Nucleosynthesis and neutrino transport calculations are carried out during the hydrodynamic calculation. The nucleosynthesis {is} confirmed later with more detailed calculations using a much larger nuclear network. 

Our main findings are: 
\begin{itemize} 
\item Bare collapse  forms a light neutron star {within the observed range of mass for NSs in BNS systems.} When this collapse takes place in a the binary, due to the small mass ejected the binary remains bound, in almost circular orbit, and will have a small kick velocity. 
\item The collapse ejects $\sim 0.1\solarmass$ of iron group elements with an outer shell composed mostly of \Ni56. 
Typical velocities of the ejected shell are $2$-$3 \times 10^9$cm/sec. 
{  \item The typical \Ni56 mass in the ejecta is $\sim 0.02\solarmass$. This result is consistent with previous explosive nucleosynthesis simulations (e.g.~\cite{moriya_2017,sawada_2021}).}
\item For Chandrasekhar mass progenitors, This radioactive ejecta results in a fast transient whose peak absolute magnitude is up to $\approx -17.2$  and whose rise time is as short as $\approx 1$ day.
\item The addition of a small envelope around the core can increase proportionally the  $^{56}$Ni mass in the ejecta. This could create even  brighter transients. { The progenitor models used for this finding } {  are somewhat artificial. However, they demonstrate the possibility of producing more \Ni56 than estimated in earlier studies. } 
\item{The kinetic energy of the ejecta, $\sim 10^{51}$ erg  can power even brighter events if the progenitor ejected a significant wind a few days prior to the collapse. }
\item{The resulting SNR would have an earlier and brighter peak than a regular SNR. However, at a later stage it will resemble a regular SNR. }
\end{itemize}

Our results demonstrate that bare collapse is a valid  mechanism for neutron star formation and in particular for the formation of  neutron star binaries. They also demonstrate that these events can be observed as fast bright transients. With transient searches getting better and better sky and temporal coverage it will be interesting to compare the BNS and fast transients statistics to see if the two are consistent.

{ Data Availability}

The data underlying this article will be shared on reasonable request to the corresponding author.

We thank Iair Arcavi,  Assaf Horesh, Samuel Jones and  Roni Waldman  for helpful remarks.
The research was supported by an advanced ERC grant TReX. 


\newpage
\appendix
\section{The effect of neutrinos}\label{appendix:A}
As discussed in~\S\ref{subsection:previous simulations}, there is a common understanding that there are three main mechanisms for mass ejection in ECSNe; the \textit{prompt}, the \textit{delayed-neutrino}, and the \textit{neutrino-driven wind} mechanisms.
There has been extensive research comparing between the first two; Various different 1D simulations have shown that when neutrino physics is included, the bounce shock is stalled and a prompt explosion fails~\citep{fryer_1999,baron_1987a,baron_1987b,mayle_1988,hillebrandt_1984,kituara_2006}. In most cases, neutrino heating was able to revive the shock on a time scale of $\sim 100$ ms, comparable to the neutrino diffusion time scales. Consequently, there is a delay by a similar amount between the time of bounce and the time of shock breakout. It is suggested that this time delay should be between $20$-$200$ ms~\citep{fryer_1999}.
Other simulations~\citep{fryer_1999,sharon_kushnir_2020} showed that when neutrino physics is not included, a prompt explosion succeeds and shock breakout occurs only a few ms after bounce. 

In this short appendix we revisit this topic.
We ran two identical simulations of our isentropic Chandrasekhar mass progenitor, once when neutrino transport was enabled and once when it was turned off. In the latter case, the electron fraction remains constant during the entire simulation.
Without neutrinos, the progenitor was stable in our simulations. To induce the collapse, we artificially reduced the value of the electron fraction $Y_e$ to $0.2$ for the inner $5\times 10^{-2}\solarmass$ of the progenitor, at time $0$. To allow a fair comparison, we did the same for the simulation with neutrinos enabled, even though it would have still collapsed without this adjustment.

\begin{figure}
\begin{center}
\includegraphics[scale=0.34]{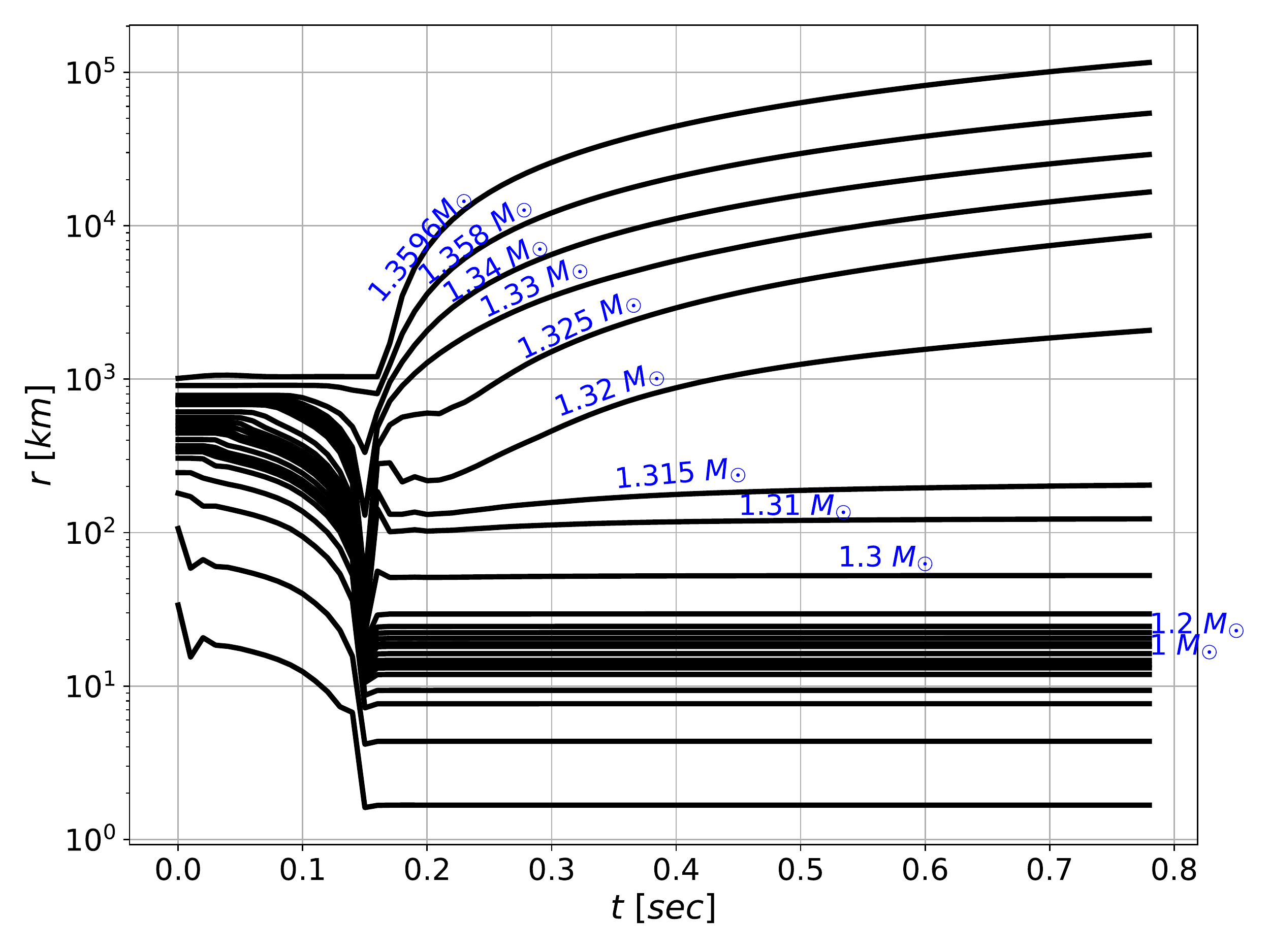}
\caption{Trajectories of  mass elements for the isentropic  progenitor, with neutrino physics turned off. The collapse is induced by reducing the value of $Y_e$ to $0.2$ at the inner $5\times 10^{-2}\solarmass$ of the progenitor, as it is otherwise stable in our simulations. 
Bounce and shock breakout occur only a few ms apart, indicative of the prompt mechanism. Only $\approx 0.05\solarmass$ is ejected, a smaller amount compared to the case with neutrinos enabled, and comparable to the amount of ejecta in the work of \protect\cite{sharon_kushnir_2020}.}
\label{fig:no neutrino}
\end{center}
\end{figure}

\begin{figure}
\begin{center}
\includegraphics[scale=0.34]{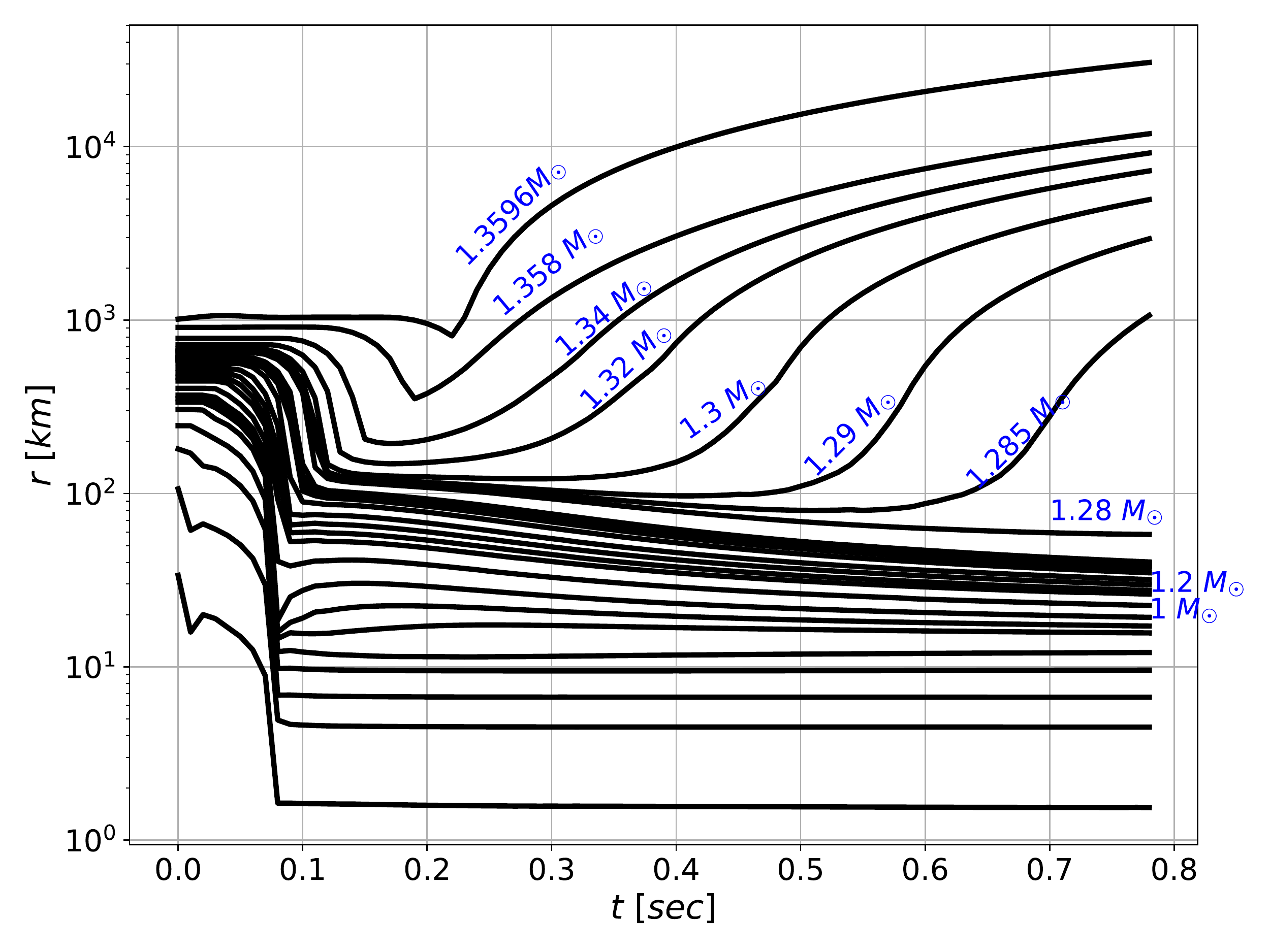}
\caption{Trajectories of  mass elements for the isentropic  progenitor, with neutrino physics enabled. The initial value of $Y_e$ was reduced to $0.2$ at the inner $5\times 10^{-2}\solarmass$ of the progenitor, as in Figure~\ref{fig:no neutrino}, for a fair comparison. Bounce occurs slightly earlier compared to the case without neutrinos, because electron capture expedites the collapse.
There is a clear delay of nearly $150$ ms between the time of bounce and shock breakout, indicative of the delayed-neutrino mechanism. The amount of delay agrees with previous works. The amount of mass ejected is larger compared to the case without neutrinos.}
\label{fig:with neutrino}
\end{center}
\end{figure}

Trajectories of mass elements as a function of time are shown in Figures~\ref{fig:no neutrino}-~\ref{fig:with neutrino} for the cases without and with neutrinos, respectively. In the case without neutrinos shock breakout occurs only a few ms after bounce, indicative of the prompt mechanism. Contrary, in the case with neutrinos there is a clear delay of nearly $150$ ms between bounce and the ejection of the outermost layer, indicative of the delayed neutrino mechanism. The amount of delay agrees with previous works.
The inclusion of neutrinos increases the amount of ejecta\footnote{Note that in the simulations shown in this appendix the ejecta mass is smaller than the ejecta mass obtained in~\S\ref{sec:results}. This occurs due to the fact that here the initial progenitor is modified to have lower $Y_e$ in the center at the beginning of the simulation.}, which could be a part of the reason why some previous works report less ejecta than us, c.f.~\cite{sharon_kushnir_2020}.

\end{document}